\documentclass{article}
\usepackage{amsmath}
\usepackage{DejaVuSansCondensed}
\usepackage{multirow}
\usepackage{arydshln}
\usepackage{color, colortbl}
\usepackage[table]{xcolor}
\usepackage{hhline}
\usepackage[T1]{fontenc}
\usepackage{url}
\usepackage[round, sort&compress, numbers]{natbib}
\usepackage{setspace}
\usepackage[utf8]{inputenc} 
\usepackage{graphicx}
\usepackage[percent]{overpic}
\usepackage{authblk}

\definecolor{LightCyan}{rgb}{0.88,1,1}
\definecolor{Maroon}{cmyk}{0,0.87,0.68,0.32}
\title{Neutral bots probe political bias\\on social media}

\author[1]{Wen Chen}
\author[1,2]{Diogo Pacheco}
\author[1]{\\ Kai-Cheng Yang}
\author[1]{Filippo Menczer\thanks{Corresponding author. Email: fil@iu.edu}}
\affil[1]{Observatory on Social Media, Indiana University, Bloomington, USA}
\affil[2]{Department of Computer Science, University of Exeter, UK}

\date{}

\begin{document}
\maketitle

\newcommand{\drifter}{drifter}
\newcommand{\Drifter}{Drifter}
\newcommand{\drifters}{drifters}
\newcommand{\Drifters}{Drifters}
\newcommand{\mention}[1]{{\texttt{@#1}}}
\newcommand{\SI}{Supplementary Information}

\begin{abstract} 
Social media platforms attempting to curb abuse and misinformation have been accused of political bias. We deploy neutral social bots who start following different news sources on Twitter, and track them to probe distinct biases emerging from platform mechanisms versus user interactions. We find no strong or consistent evidence of political bias in the news feed. Despite this, the news and information to which U.S. Twitter users are exposed depend strongly on the political leaning of their early connections. The interactions of conservative accounts are skewed toward the right, whereas liberal accounts are exposed to moderate content shifting their experience toward the political center. Partisan accounts, especially conservative ones, tend to receive more followers and follow more automated accounts. Conservative accounts also find themselves in denser communities and are exposed to more low-credibility content. 
\end{abstract}

\section*{Introduction}

Compared with traditional media, online social media can connect more people in a cheaper and faster way than ever before.
As a large portion of the population frequently use social media to generate content, consume information, and interact with others~\cite{perrin2019share}, online platforms are also shaping the norms and behaviors of their users.
Experiments show that simply altering the messages appearing on social feeds can affect the online expressions and real-world actions of users~\cite{kramer2014experimental,bond201261}, and that social media users are sensitive to early social influence~\cite{muchnik2013social,weninger2015random}.
At the same time, discussions on social media tend to be polarized around critical yet controversial topics like elections~\cite{conover2011political,Conover2012,hanna2013partisan},
vaccination~\cite{schmidt2018polarization}, and climate change~\cite{williams2015network}.
Polarization is often accompanied by the segregation of users with incongruent views into so-called echo chambers~\cite{Jamieson2008,garrett2009echo,lee2014social,flaxman2016filter,sunstein2017republic,garimella2018political}, homogeneous online communities that have been associated with ideology radicalization and misinformation spreading~\cite{wojcieszak2010don,del2016spreading,Bright2016,Nikolov2020partisanship}.

Countering such undesirable phenomena requires a deep understanding of their underlying mechanisms.
On the one hand, online vulnerabilities have been associated with several socio-cognitive biases of humans~\cite{del2017modeling,sasahara2019inevitability,HillsProliferation18}, including selection of belief-consistent information~\cite{nickerson1998confirmation} and the tendency to seek homophily in social ties~\cite{mcpherson2001birds}.
On the other hand, web platforms have their own algorithmic biases~\cite{nikolov2015measuring,baeza2018bias,Nikolov2018biases}.
For example, ranking algorithms favor popular and engaging content, which may create a vicious cycle amplifying noise over quality~\cite{ciampaglia2018algorithmic}. Exposure to engagement metrics may also increase vulnerability to misinformation~\cite{Fakey2020}. 
For a more extreme illustration, recent studies and media reports suggest that the YouTube recommendation system might lead to videos with more misinformation or extreme viewpoints regardless of the starting point~\cite{ribeiro2020auditing}.

Beyond the socio-cognitive biases of individual users and algorithmic biases of technology platforms, we have a very limited understanding of how collective interactions mediated by social media may bias the view of the world that we obtain through the online information ecosystem. 
The major obstacle is the complexity of the system --- not only do users exchange huge amounts of information with large numbers of others via many hidden mechanisms, but these interactions can be manipulated overtly and covertly by legitimate influencers as well as inauthentic, adversarial actors who are motivated to influence opinions or radicalize behaviors~\cite{thompson2011radicalization}. Evidence suggests that malicious entities like social bots and trolls have already been deployed to spread misinformation and influence public opinion on critical matters~\cite{shao2018spread,stella2018bots,broniatowski2018weaponized,zannettou2019disinformation,Caldarelli2020bots}.

In this study, we aim to reveal biases in the news and information to which people are exposed in social media ecosystems. 
We are particularly interested in clarifying the role of social media interactions during the polarization process and the formation of echo chambers.
We therefore focus on U.S. political discourse on Twitter since this platform plays an important role in American politics and strong polarization and echo chambers have been observed~\cite{conover2011political,colleoni2014echo}. 
Twitter forms a directed social network where an edge from a \textit{friend} node to a \textit{follower} node indicates that content posted by the friend appears on the news feed of the follower.

Our goal is to study ecosystem bias, which includes both potential platform bias and the net effects of interactions with users of the social network (organic or not) that are mediated by the platform's mechanisms and regulated by its policies. 
While we only attempt to separate platform effects from naturally occurring biases in the narrow case of the feed curation, our investigation targets the overall bias experienced by users of the platform. 
This requires the exclusion of biases from individual users, which is a challenge when using traditional observational methods --- it would be impossible to separate ecosystem effects from confounding factors that might affect the actions of tracked human accounts, such as age, gender, race, ideology, and socioeconomic status. 
We thus turn to a method that removes the need to control for such confounding factors by leveraging social media accounts that mimic human users but are completely controlled by  algorithms, known as \emph{social bots}~\cite{hargreaves2019fairness}. 
Here we deploy social bots with neutral (unbiased) and random behavior as instruments to probe exposure biases in social media.
We call our bots \emph{\drifters{}} to distinguish their neutral behavior from other types of benign and malicious social bots on Twitter~\cite{ferrara2016rise}. 

\Drifters{} are designed with an identical behavior model but with the only distinctive difference of their initial friend --- the very first account they follow. After this initial action that represents the single independent variable (treatment) in our experiment, each \drifter{} is let loose in the wild. To be sure, while all \drifters{} have identical behaviors, their actions are different and depend on their initial conditions. We expect that a \drifter{} who starts following liberal accounts will be more likely to be exposed to liberal content, to share some of this content, to be followed by liberal accounts, and so on. But these actions are driven by platform mechanisms and social interactions, not by political bias in the treatment: the behavioral model has no way of distinguishing between liberal, conservative, or any type of content. The \drifter{} actions are therefore part of the dependent variables (outcomes) measured by our experiment.

This methodology allows us to examine the combined biases that stem both from Twitter's system design and algorithms, and from the organic and inorganic social interactions between the \drifters{} and other accounts. Our research questions are: (i)~How are influence and exposure to inauthentic accounts, political echo chambers, and misinformation impacted by early actions on a social media platform? And (ii)~Can such differences be attributed to political bias in the platform's news feed? 

To answer these questions, we initialized \drifters{} from news sources across the political spectrum. After five months, we examined the content consumed and generated by the \drifters{} and analyzed (i)~characteristics of their friends and followers, including their liberal-conservative political alignment inferred by shared links and hashtags; (ii)~automated activity measured via machine learning methods; and (iii)~exposure to information from low-credibility sources identified by news and fact-checking organizations. 

We find that the political alignment of the initial friend has a major impact on the popularity, social network structure, exposure to bots and low-credibility sources, and political alignment manifested in the actions of each \drifter{}. However, we find no evidence that these outcomes can be attributed to platform bias. 
The insights provided by our study into the political currents of Twitter's information ecosystem can aid the public debate about how social media platforms shape people's exposure to political information. 

\section*{Results}

All \drifters{} in our experiment follow the same behavior model, whose design is intended to be neutral, not necessarily realistic. Each \drifter{} is activated at random times to performs actions. Action types, such as  tweets, likes, and replies, are selected at random according to predefined probabilities. For each action, the model specifies how to select a random target, such as a tweet to be retweeted or a friend to be unfollowed. Time intervals between actions are drawn from a broad distribution to produce realistic bursty behaviors. See Methods for further details.

We developed 15 \drifter{} bots, divided them into five groups, and initialized each \drifter{} in the same group with the same initial friend. Each Twitter account used as a first friend a popular news source aligned with the left, center-left, center, center-right, or right of the U.S. political spectrum (see details in Methods).
We refer to the \drifters{} by the political alignment of their initial friends; for example, bots initialized with center-left sources are called ``C.~Left'' \drifters{}.

Between their deployment on July 10, 2019 and until their deactivation on December 1, 2019, we monitored the behaviors of the \drifters{} and collected data on a daily basis.
In particular, we measured: (1)~the number of followers of each \drifter{} to compare their ability to gain influence; (2)~the echo-chamber exposure of each \drifter{}; (3)~the likely automated activities of friends and followers of the \drifters{}; (4)~the proportion of low-credibility information to which the \drifters{} are exposed; and (5)~the political alignment of content generated by the \drifters{} and their friends to probe political biases.

\subsection*{Influence}

The number of followers can be used as a crude proxy for influence~\cite{cha2010measuring}. To gauge how political alignment affects influence dynamics, Fig.~\ref{fig:followers} plots the average number of followers of \drifters{} in different groups over time. 
To compare the growth rates of different groups, we considered consecutive observations of the follower counts of each \drifter{} and aggregated them across each group ($n=387$ for Left, 373 for C.~Left, 389 for Center, 387 for C.~Right, and 386 for Right).  
Two trends emerged from t-tests (all t-tests in this and the following analyses are two-sided).
First, \drifters{} with the most partisan sources as initial friends tend to attract significantly more followers than center \drifters{} ($d.f.=774, t=5.13, p<0.001$ for Left vs. Center and $d.f.=773, t=8.00, p<0.001$ for Right vs. Center). 
Second, \drifters{} with right-leaning initial sources gain followers at a significantly higher rate than those with left-leaning initial sources ($d.f.=771, t=3.84, p<0.001$ for Right vs. Left).
More details about this and additional statistical tests are shown in Supplementary Notes, confirming the robustness of these findings.

The differences in influence among \drifters{} could be affected not only by the political alignment, but also by other characteristics of their initial friends. To disentangle these factors, we measured the correlation between the number of \drifter{} followers and two features of their initial friends: their overall influence and their popularity among other politically aligned accounts. While \drifter{} influence is not affected by the overall influence of the initial friends, it is positively correlated with their popularity among politically aligned accounts (see Supplementary Notes). This is consistent with evidence that users with shared partisanship are more likely to form social ties~\cite{Moslehe2022761118}, as we explore next. 

\subsection*{Echo chambers}

We define echo chambers as dense and highly clustered social media neighborhoods that amplify exposure to homogeneous content. 
To investigate whether the \drifter{} bots find themselves in such echo chambers, let us consider the ego network of each \drifter{}, i.e., the network composed by the \drifter{} and its friends and followers. 
We can use density and transitivity of ego networks as proxies for the presence of echo chambers. 
Density is the fraction of node pairs that are connected in a network. 
Transitivity measures the fraction of possible triangles that are actually present among the nodes. High transitivity means that friends and followers are likely to follow each other too. See Methods for further details. 

\begin{figure}[!ht]
    \centering
    \includegraphics[width=\textwidth]{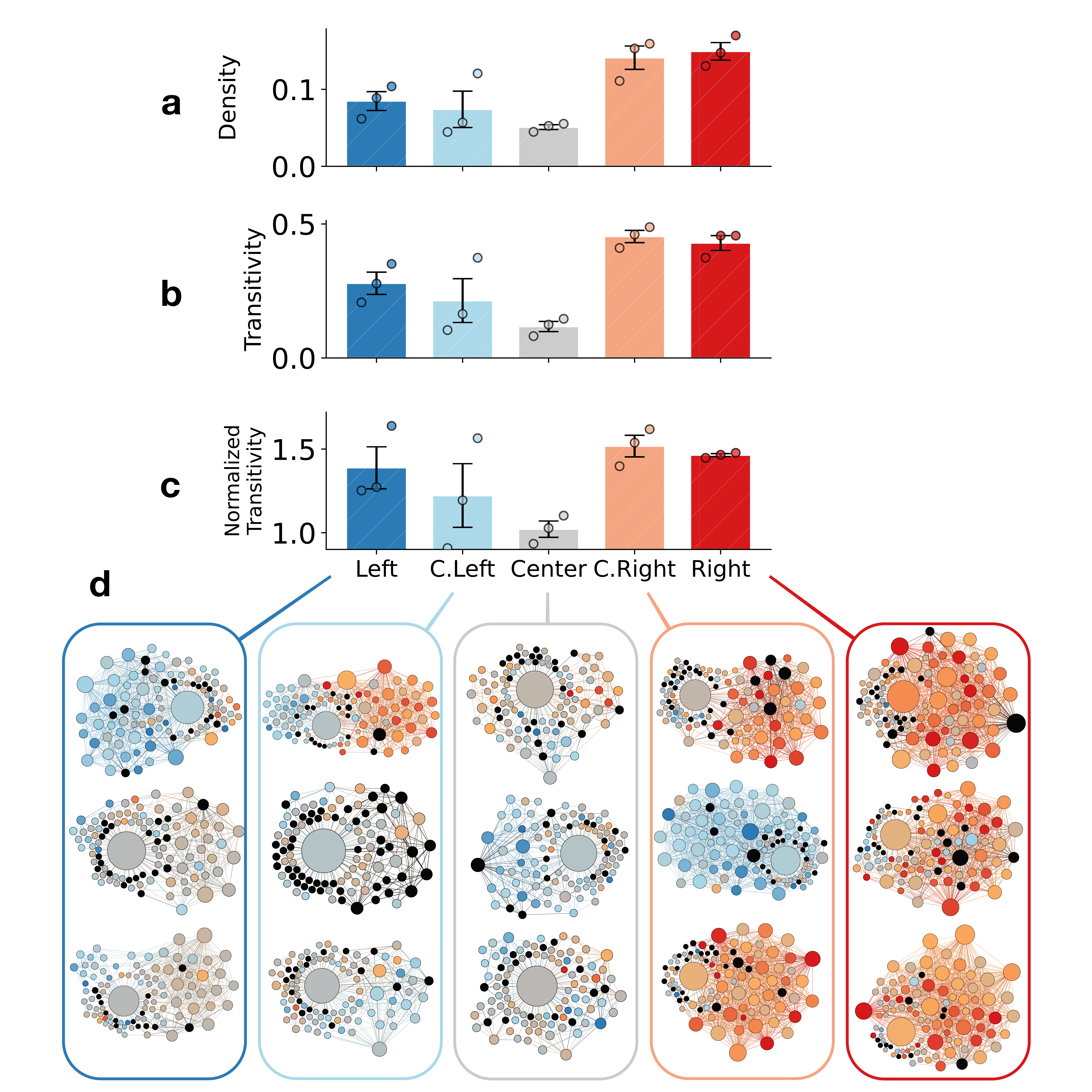}
    \caption{\textbf{Echo chamber structure around \drifters{}.} \textbf{a}~Density, \textbf{b}~transitivity, and \textbf{c}~normalized transitivity of \drifters{} ego networks in different groups. 
    Error bars indicate standard errors ($n=3$ \drifters{} in each group).
    \textbf{d}~Ego networks of the \drifters{} in the five groups. Nodes represent accounts and edges represent friend/follower relations. Node size and color represent degree (number of neighbors) and political alignment of shared links, respectively. Black nodes have missing alignment score due to not sharing political content. Source data are provided as a Source Data file.
    }
    \label{fig:clus_coeff}
\end{figure}

Fig.~\ref{fig:clus_coeff}\textbf{a--b} shows the average density and transitivity of ego networks for the \drifters{} (see details in Methods). 
Since the two metrics are correlated in an ego network, Fig.~\ref{fig:clus_coeff}\textbf{c} also plots the transitivity rescaled by that of shuffled random networks (see Methods). 
The ego networks of Right \drifters{} are more dense than those of Center \drifters{} ($d.f.=4, t=-8.28, p=0.001$), whereas the difference in density is not significant between Center and Left \drifters{} ($d.f.=4, t=-2.68, p=0.055$).
Right account networks also have higher transitivity than Center networks ($d.f.=4, t=-9.31, p<0.001$); the difference for Left vs. Center is less significant ($d.f.=4, t=-3.53, p=0.024$). 
Right accounts are more clustered than centrists even when accounting for the difference in density ($d.f.=4, t=-8.96, p<0.001$) while the difference is not significant for Left vs. Center ($d.f.=4, t=-2.73, p=0.053$).
Furthermore, Right \drifters{} are in stronger echo chambers than Left \drifters{} ($d.f.=4, t=-3.84, p=0.019$ for density and $d.f.=4, t=-3.02, p=0.039$ for transitivity). However, the difference in normalized transitivity between Left and Right is not significant ($d.f.=4, t=-0.60, p=0.579$), indicating that the higher clustering on the right is explained by the density of social connections.

To get a better sense of what these echo chambers look like, Fig.~\ref{fig:clus_coeff}\textbf{d} maps the ego networks of the 15 \drifters{}. In addition to the clustered structure, we observe a degree of homogeneity in shared content as illustrated by the colors of the nodes, which represent the political alignment of the links shared by the corresponding accounts (see Methods; similar results are obtained by measuring political alignment based on shared hashtags). In general, the neighbors of a \drifter{} tend to share links to sources that are politically aligned with the \drifter{}'s first friend. We note a few exceptions, however. The Left \drifters{} and their neighbors are more moderate, having shifted their alignment toward the center. One of the C.~Left \drifters{} has become connected to many conservative accounts, shifting its alignment to the right. And one of the C.~Right \drifters{} has shifted its alignment to the left, becoming connected to mostly liberal accounts after randomly following \mention{CNN}, a left-leaning news organization. In most cases, \drifters{} find themselves in structural echo chambers where they are exposed to content with homogeneous political alignment that mirrors their own. 

\subsection*{Automated activities}

\begin{figure}[!ht]
    \centering
    \includegraphics[width=\columnwidth]{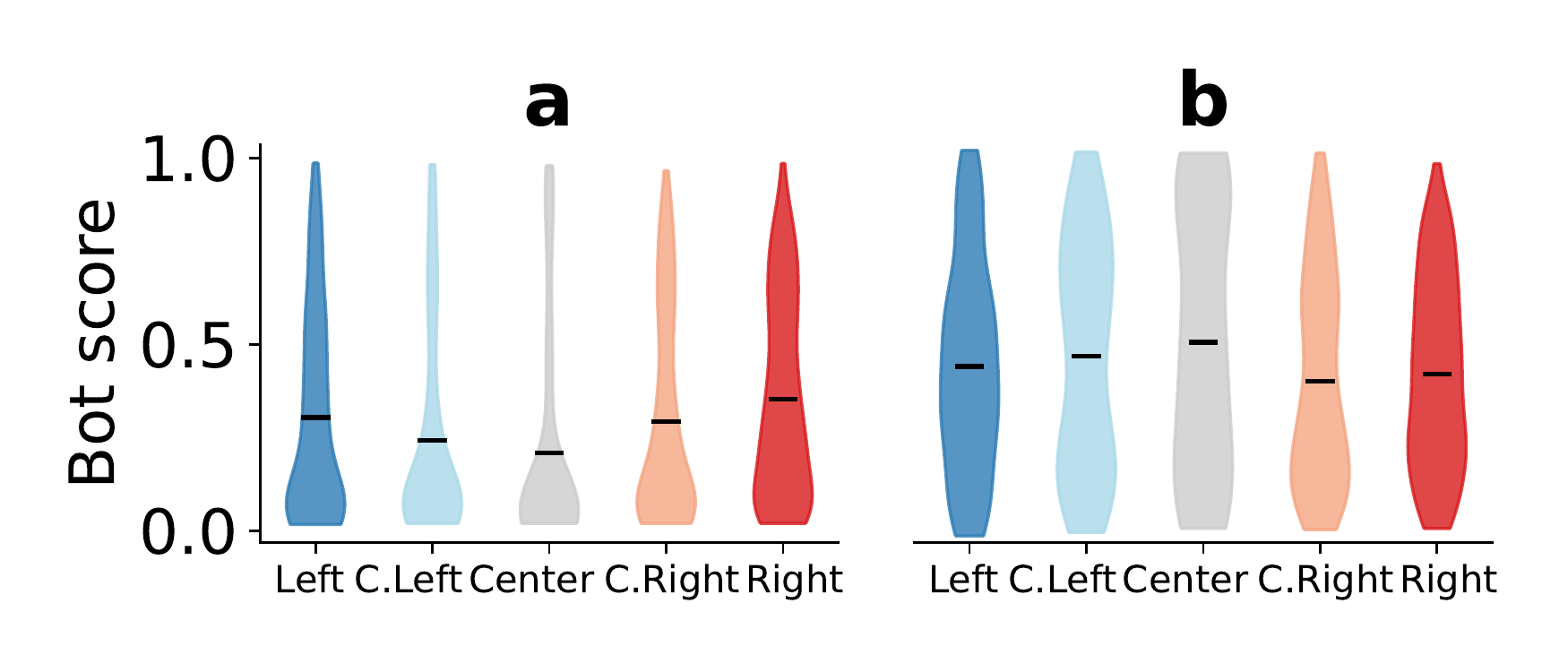}
    \caption{\textbf{Distributions of bot scores of friends and followers of \drifters{}.} The bot score is a number between zero and one, with higher scores signaling likely automation. For each group, we consider the union of (\textbf{a}) friends and (\textbf{b}) followers of the \drifters{} in that group. Bars indicate averages. For friends, $n=$ 282 (Left), 261 (C.~Left), 206 (Center), 323 (C.~Right), 414 (Right). For followers, $n=$ 172 (Left), 118 (C.~Left), 65 (Center), 205 (C.~Right), 299 (Right). Source data are provided as a Source Data file.}
    \label{fig:connections_botscore-bots}
\end{figure}

Automated accounts known as social bots were actively involved in online discussions about recent U.S. elections \cite{bessi2016social,deb2019perils,shao2018spread}. 
It is therefore expected for the \drifters{} to encounter bot accounts. 
We used the Botometer service~\cite{varol2017online,yang2019arming} to collect bot scores of friends and followers of the \drifters{}. 
We report the average bot scores for both friends and followers of the \drifters{} in Fig.~\ref{fig:connections_botscore-bots}.
Unsurprisingly, \drifters{} are more likely to have bots among their followers than among their friends, across the political spectrum. 
Focusing on the friends reveals a more serious potential vulnerability of social media users. We find that accounts followed by partisan \drifters{} are more bot-like than those followed by centrist \drifters{} ($d.f.=618, t=-6.14, p<0.001$ for Right vs. Center and $d.f.=486, t=-3.67, p<0.001$ for Left vs. Center). 
Comparing partisans and moderates, Right \drifters{} follow accounts that are more bot-like than C.~Right \drifters{} ($d.f.=735, t=-3.01, p=0.003$), while the difference is smaller on the liberal side ($d.f.=541, t=-2.56, p=0.011$ for Left vs. C.~Left). 
Among partisans, Right \drifters{} follow accounts that are slightly more bot-like than Left ones ($d.f.=694, t=-2.33, p=0.020$).

\subsection*{Exposure to low-credibility content}

\begin{figure}[!ht]
    \centering
    \includegraphics[width=0.7\textwidth]{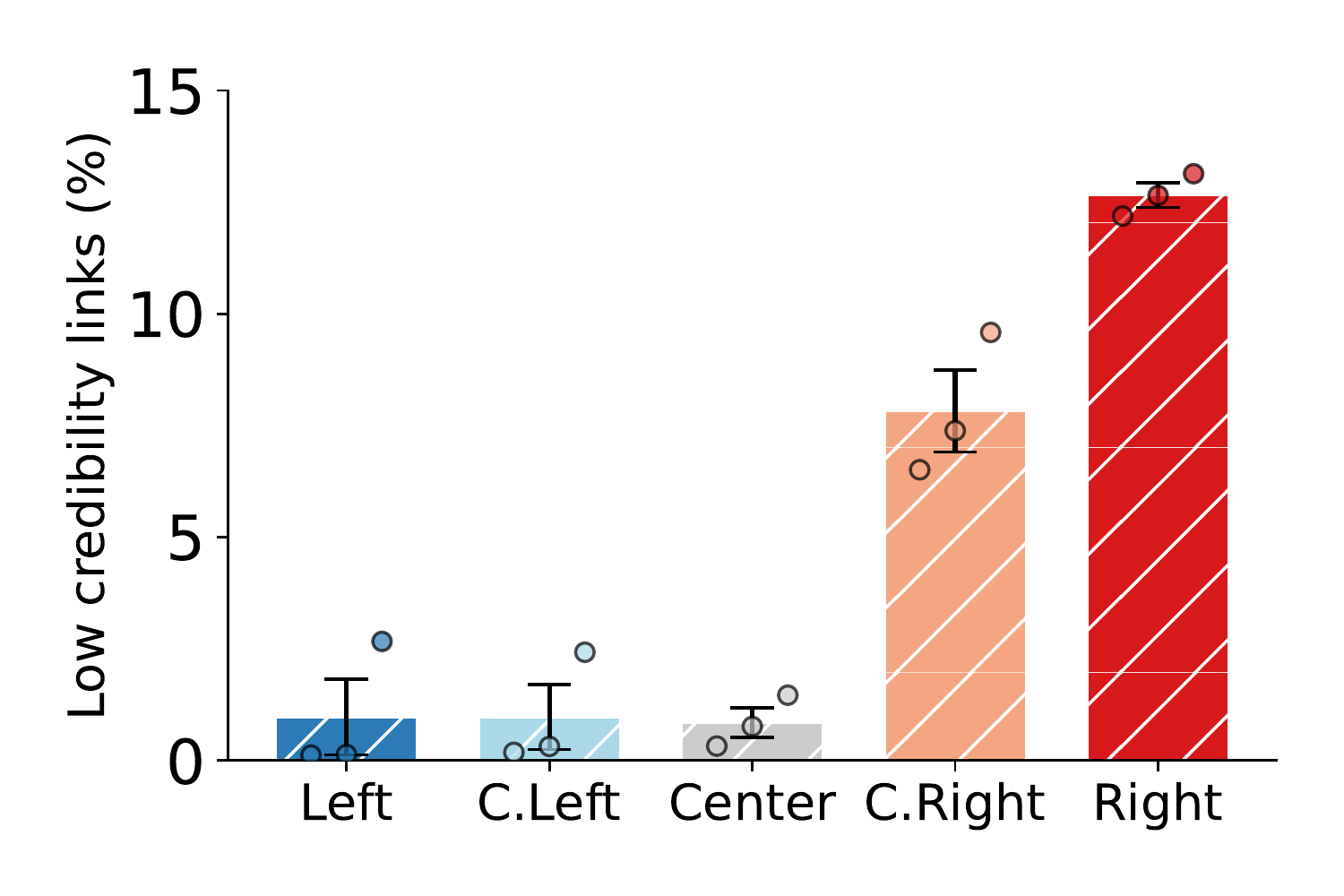}
    \caption{\textbf{Exposure to low-credibility content.} The bars express the proportions of low-credibility links in the home timelines of different \drifter{} groups. Error bars indicate standard errors ($n=3$ \drifters{} in each group). Source data are provided as a Source Data file.}
    \label{fig:low_credibility}
\end{figure}

Since the 2016 U.S. presidential election, concern has been heightened about the spread of misinformation in social media~\cite{Lazer-fake-news-2018}. 
We consider a list of low-credibility sources that are known to publish false and misleading news reports, conspiracy theories, junk science, and other types of misinformation (details about low-credibility sources are found in Methods).
We analyze exposure to content from these low-credibility sources for different groups of \drifters{} in Fig.~\ref{fig:low_credibility}. 
We observe that Right \drifters{} receive more low-credibility content in their news feeds than other groups ($t=5.06, p=0.007$ compared to C.~Right; $p<0.001$ for other groups: $t=27.47$ compared to Center, $t=15.06$ to C.~Left, and $t=13.14$ to Left; $d.f.=4$ in all cases).
Almost 15\% of the links that appear in the timelines of Right \drifters{} are from low-credibility sources.
We also measured the absolute number of low-credibility links, and used the total number of tweets or the number of tweets with links as the denominator of the proportions; the same pattern emerges in all cases.

We used \mention{BreitbartNews} as the initial friend account for Right \drifters{} because it is one of the most popular conservative news sources. 
Although \textit{Breitbart News} appears in lists of hyper-partisan sources used in the literature, to prevent biasing our results, \textit{Breitbart News} is not labeled as a low-credibility source in this analysis and does not contribute to the proportions in Fig.~\ref{fig:low_credibility}.

\subsection*{Political alignment and news feed bias}
\label{sec:valence}

We wish to measure the political alignment of content consumed and produced by \drifters{}. Given a link (URL), we can extract the source (website) domain name and obtain an alignment score based on its known political bias. Similarly, given a hashtag, we can calculate a score based on a computational technique that captures the political alignment of different hashtags. These scores can then be averaged across the links or hashtags contained in a feed of tweets to measure their aggregate political alignment. Further details can be found in Methods. 

The \textit{home timeline} (also known as news feed) is the set of tweets to which accounts are exposed. The \textit{user timeline} is the set of tweets produced by an account. In Fig.~\ref{fig:timeline_valence}\textbf{a,b,d,e} we observe how the political alignment of information to which \drifters{} are exposed in their home timelines ($s_h$) and of content generated by them in their user timelines ($s_u$) changed during the experiment. 
The initial friends strongly affect the political trajectories of the \drifters{}.
Both in terms of information to which they are exposed and content they produce, \drifters{} initialized with right-leaning sources stay on the conservative side of the political spectrum. Those initialized with left-leaning sources, on the other hand, tend to drift toward the political center: they are exposed to more conservative content and even start spreading it. These findings are robust with respect to the method used to calculate political alignment, whether  based on hashtags (Fig.~\ref{fig:timeline_valence}\textbf{a,b}) or links (Fig.~\ref{fig:timeline_valence}\textbf{d,e}).

We measure the political bias of the news feed by calculating the difference between the alignment score of tweets posted by friends of the \drifters{}, $s_f$, and the score of tweets in the home timeline, $s_h$. The results are shown in Fig.~\ref{fig:timeline_valence}\textbf{c,f} for alignment computed from hashtags and links, respectively. In the case of hashtags, we observe little evidence of political bias by the news feed. For right-leaning \drifters{}, there is a small shift toward the center, suggesting a weak bias of the news feed (Fig.~\ref{fig:timeline_valence}\textbf{c}). To confirm this visual observation, we performed a paired $t$-test comparing the daily averages of home timeline scores $s_h$ and friend user timeline scores $s_f$ (Supplementary Table~3). 
The effect is small for all but the right group of \drifters{} ($p<0.001$, Cohen's $d=0.56$). 
Similarly, in the case of links (Fig.~\ref{fig:timeline_valence}\textbf{f}), we observe left bias for \drifters{} in the center group ($p<0.001$, Cohen's $d=0.76$). For the other groups, the effect is small (Supplementary Table~3). 

Further details on the bias analysis can be found in Supplementary Notes, together with trajectories of individual \drifters{}, data on follow-back rates, and descriptive statistics of the \drifters{}. 

\section*{Discussion}

The present results suggest that early choices about which sources to follow impact the experiences of social media users. This is consistent with previous studies~\cite{muchnik2013social,weninger2015random}.
But beyond those initial actions, \drifter{} behaviors are designed to be neutral with respect to partisan content and users. Therefore the partisan-dependent differences in their experiences and actions can be attributed to their interactions with users and information mediated by the social media platform --- they reflect biases of the online information ecosystem. 

\Drifters{} with right-wing initial friends are gradually embedded into dense and homogeneous networks where they are constantly exposed to right-leaning content. They even start to spread right-leaning content themselves.    
Such online feedback loops reinforcing group identity may lead to radicalization~\cite{wojcieszak2010don}, especially in conjunction with social and cognitive biases like in-/out-group bias and group polarization. 
The social network communities of the other \drifters{} are less dense and partisan.

We selected popular news sources across the political spectrum as initial friends of the \drifters{}. There are several possible confounding factors stemming from our choice of these accounts: their influence as measured by the number of followers, their popularity among users with similar ideology, their activity in terms of tweets, and so on. For example, \mention{FoxNews} was popular but inactive at the time of the experiment. Furthermore, these quantities vary greatly both within and across ideological groups (see Supplementary Methods). While it is impossible to control for all of these factors with a limited number of \drifters{}, we checked for a few possible confounding factors. We did not find a significant correlation between initial friend influence or popularity measures and \drifter{} ego network transitivity. We also found that the influence of an initial friend is not correlated with \drifter{} influence. However, the popularity of an initial friend among sources with similar political bias is a confounding factor for \drifter{} influence. Online influence is therefore affected by the echo-chamber characteristics of the social network, which are correlated with partisanship, especially on the political right~\cite{benkler2018network, Nikolov2020partisanship}. In summary, \drifters{} following more partisan news sources receive more politically aligned followers, becoming embedded in denser echo chambers and gaining influence within those partisan communities. 

The fact that right-leaning \drifters{} are exposed to considerably more low-credibility content than other groups is in line with previous findings that conservative users are more likely to engage with misinformation on social media~\cite{grinberg2019fake, Nikolov2020partisanship}.
Our experiment suggests that the ecosystem can lead completely unbiased agents to this condition, therefore it is not necessary to impute the vulnerability to characteristics of individual social media users.
Other mechanisms that may contribute to the exposure to low-credibility content observed for the \drifters{} initialized with right-leaning sources involve the actions of neighbor accounts (friends and followers) in the right-leaning groups, including inauthentic accounts that target these groups.

Although \textit{Breitbart News} was not labeled as a low-credibility source in our analysis, our finding might still be biased in reflecting this source's low credibility in addition to its partisan nature. However, \mention{BreitbartNews} is one of the most popular conservative news sources on Twitter (see Supplementary Table~1). While further experiments may corroborate our findings using alternative sources as initial friends, attempting to factor out the correlation between conservative leanings and vulnerability to misinformation~\cite{grinberg2019fake,Nikolov2020partisanship} may yield a less-representative sample of politically active accounts. 

While most \drifters{} are embedded in clustered and homogeneous network communities, the echo chambers of conservative accounts grow especially dense and include a larger portion of politically active accounts. 
Social bots also seem to play an important role in the partisan social networks; the \drifters{}, especially right-leaning ones, end up following a lot of them. Since bots also amplify the spread of low-credibility news~\cite{shao2018spread}, this may help explain the prevalent exposure of right-leaning \drifters{} to low-credibility sources. 
\Drifters{} initialized with far-left sources do gain more followers and follow more bots compared with the center group. 
However this occurs in a way that is less emphatic and vulnerable to low-credibility content compared to the right and center-right groups. Nevertheless, our results are consistent with findings that partisanship on both sides of the political spectrum increases vulnerability to manipulation by social bots~\cite{Yan2020partisanbots}.

Twitter has been accused of favoring liberal content and users. 
We examined the possible bias in Twitter's news feed, i.e., whether the content to which a user is exposed in the home timeline is selected in a way that amplifies or suppresses certain political content produced by friends.
Our results suggest this is not the case: in general, the \drifters{} receive content that is closely aligned with whatever their friends produce. 
A limitation of this analysis is that it is based on limited sets of recent tweets from  \drifter{} home timelines (see Methods). 
The exact posts to which Twitter users are exposed in their news feeds might differ due to the recommendation algorithm, which is not available via Twitter's programmatic interface.

Despite the lack of evidence of  political bias in the news feed, \drifters{} that start with left-leaning sources shift toward the right during the course of the experiment, sharing and being exposed to more moderate content. \Drifters{} that start with right-leaning sources do not experience a similar exposure to moderate information and produce increasingly partisan content. 
These results are consistent with observations that right-leaning bots do a better job at influencing users~\cite{luceri2019red}. 

In summary, our experiment demonstrates that even if a platform has no partisan bias, the social networks and activities of its users may still create an environment in which unbiased agents end up in echo chambers with constant exposure to partisan, inauthentic, and misleading content. 
In addition, we observe a net bias whereby the \drifters{} are drawn toward the political right. On the conservative side, they tend to receive more followers, find themselves in denser communities, follow more automated accounts, and are exposed to more low-credibility content. 
Users have to make extra efforts to moderate the content they consume and the social ties they form in order to counter these currents and create a healthy and balanced online experience.

Given the political neutrality of the news feed curation, we find no evidence for attributing the conservative bias of the information ecosystem to intentional interference by the platform.
The bias can be explained by the use (and abuse) of the platform by its users, and possibly to unintended effects of the policies that govern this use: neutral algorithms do not necessarily yield neutral outcomes. 
For example, Twitter may remove or demote information from low-credibility sources and/or inauthentic accounts, or suspend accounts that violate its terms. To the extent that such content or users tend to be partisan, the net result would be a bias toward the center. 
How to design mechanisms capable of mitigating emergent biases in online information ecosystems is a key question that remains open for debate.

\section*{Broader impact}

The methodology based on neutral social bots can be applied to study a number of biases of social media platforms and their information ecosystems, in addition to political bias. Applications of our method could study gender and racial bias, hate speech, and algorithmic bias. Bots could also be used for benign interventions, such as posting content from trustworthy sources in response to harmful health misinformation. However, when bots interact with humans without informed consent, even in the absence of deception, there are trade-offs between potential benefits and risks to the users. For example, our neutral bots might share misinformation or contribute to polarization. Therefore the ethical implications of such applications must be evaluated carefully.  

Our main finding --- that the information Twitter users see in their news feed depends on the political leaning of their early connections --- has societal implications. It may increase awareness among social media users about the implicit biases of their online connections and their own vulnerabilities to selective exposure of information, or worse --- influence campaigns, manipulation, misinformation, and polarization. The absence of strong or consistent evidence of political bias in the Twitter news feed will help inform the public debate about social media platform regulation. In particular, in the U.S., some critics want to revoke Section 230 of the Communications Decency Act, which protects internet companies from most lawsuits related to user-generated content. This position is based on claims that platforms censor political speech. Our findings do not support such claims.

The media have reported extensively on links between online misinformation and harmful behaviors in domains like health and elections. Our data is limited to online behaviors and cannot gauge the impact of political bias on real-world actions. Our research also does not address platform policies and their enforcement, nor other types of algorithmic bias. In particular, our study is unable to evaluate the effects of Twitter's personalized ranking of news feed posts, friend recommendations, suspension of accounts, or ads. A final limitation of our experiment is in the small number of deployed neutral bots, motivated by the ethical considerations discussed above.

Further research questions remain open concerning political bias in information ecosystems. How would the findings be affected by deploying a larger number of bots and starting from news sources with greater or smaller diversity in popularity, influence, activity, or political slant? What is the net effect of major changes in platform policy/enforcement, such as the takedown of misinformation ``superspreaders''? Or the migration of radical users to other platforms? Can the findings on Twitter be generalized to platforms with different user demographics, such as Facebook, Instagram, or TikTok, and more partisan user populations, such as Parler or Gab? Finally, our study is U.S.-centric. Similar questions could be explored in the political contexts of other countries, for example those with authoritarian governments, possibly yielding different conclusions.

\section*{Methods}

Here we provide details about the design of \drifter{} bots, the computation of political alignment metrics, the identification of low-credibility sources, and the characterization of echo chambers. All  \drifter{} activities are managed through Twitter's application programmatic interface (API).

\subsection*{\Drifter{} behavior model}
\label{sec:bot_design}

\Drifter{} bots are the key instrument for this study. They are designed to mimic social media users, so that the data collected from their actions and interactions reflects experiences on the platform. 
The \drifters{} are intended to have a neutral behavior, even if it is not necessarily realistic. 
For example, they lack any ability to comprehend the content to which they are exposed or the users with whom they interact. 
All actions are controlled by a stochastic model which was consistent across treatments and unchanged during the experiment. 

When creating the \drifter{} profiles, we avoided any political references that would bias the experiment. \Drifter{} bots also complied with Twitter terms and guidelines. Twitter does not prohibit bots as long as they do not engage in any prohibited behaviors like spam, deception, and other abuse. \Drifters{} did not engage in any such behaviors. In particular, to avoid impersonating any human, the \drifter{} accounts were named after fictional robots in the arts and literature; used robot images in the public domain for their profiles; and used random quotes as profile descriptions. In addition, the \drifters{} did not engage in direct messages nor use any advanced natural language generative models. If a human user started a conversation with any of our \drifters{}, it would be obvious to them that those accounts were bots.

Since the \drifters{} interacted with human subjects, the experimental protocol was vetted and approved by the Indiana University's ethics board. The protocol did not require disclosure to Twitter users with whom the bots interacted. However, a large number of \drifters{} could have negative impact by spreading misinformation, reinforcing echo chambers, or amplifying malicious accounts. Therefore we deployed 15 \drifters{} as a trade-off between limiting potential harm and achieving statistically significant results. 

Like many human behaviors, social media activity is bursty~\cite{ghoshentropy}. To reproduce this feature, we draw time intervals $\Delta t$ between two successive actions from a power-law distribution $P(\Delta t) \sim \Delta t^{-\alpha}$, with exponent $\alpha=0.9$ manually tuned to minimize the bot score obtained from the Botometer service.  
The distribution was cutoff at a maximum sleep duration of seven hours between consecutive actions. 
Intervals were further scaled to obtain an average frequency of 20--30 actions per day --- a typical activity level of normal users estimated from sampled active Twitter accounts.
Moreover, the \drifters{} were inactive between midnight and 7~a.m., a typical sleep cycle for social media users~\cite{Barbosa2018}. 

\begin{figure}[!ht]
    \centering
    \includegraphics[width=\linewidth]{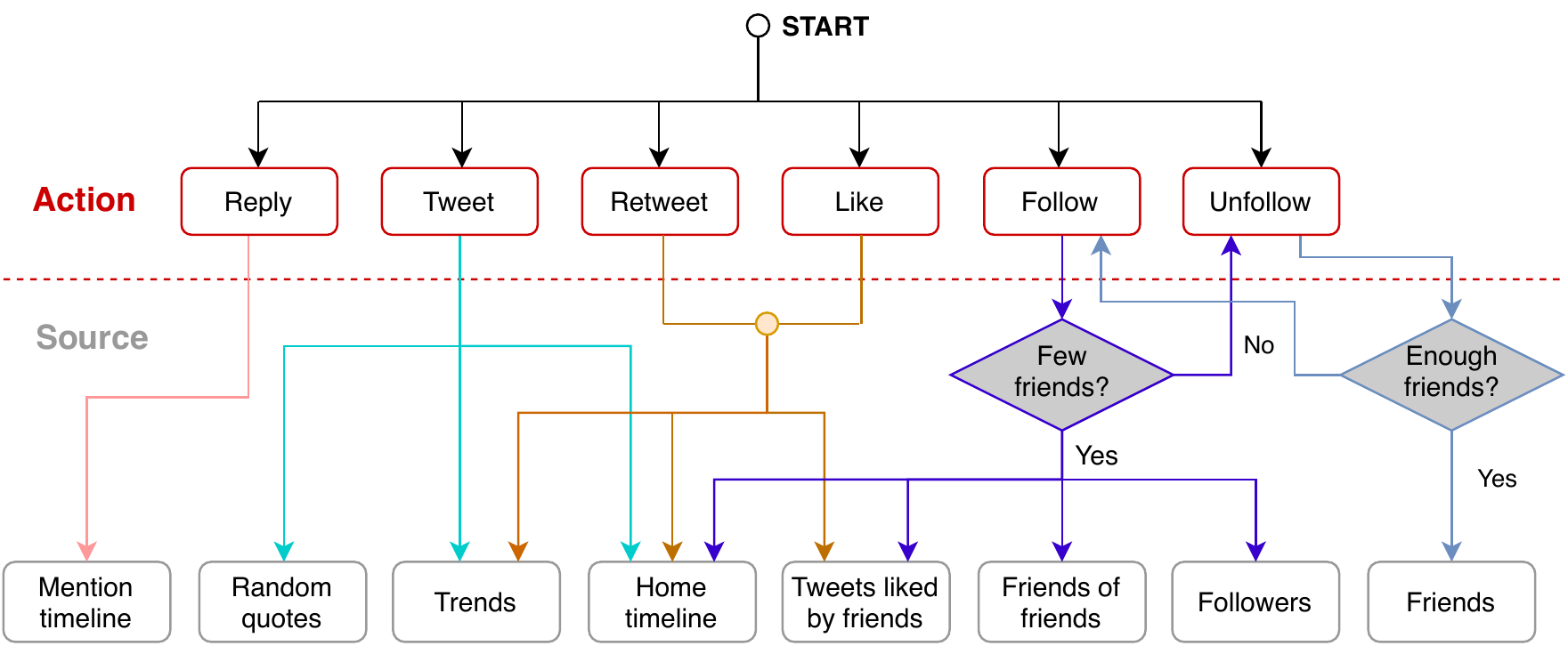}
    \caption{\textbf{\Drifter{} bot behavior model.} Each action box is connected with boxes that indicate the sources used for that action.  
    For example, the source of a retweet can be a trending tweet, a tweet in the home timeline, or a tweet liked by a friend. Links to actions and sources are associated with probabilities. 
    Follow and unfollow actions require additional constraints to be satisfied (gray diamonds).}
    \label{fig:workflow}
\end{figure}

Every time a \drifter{} is activated, it randomly selects an action and a source as illustrated in Fig.~\ref{fig:workflow}. Actions include tweets, retweets, likes, replies, etc. Sources include the home timeline, trends, friends, etc. Each action is selected with a predefined probability. Given the selected action, one of a set of possible sources is selected with a predefined conditional probability. A random object is then drawn from the source and the action is performed on it. For example, if the action is a retweet and the source is the home timeline, then a random tweet in the \drifter{}'s home timeline is retweeted. Non-English sources (users and tweets) are disregarded when they can be identified from metadata. 
Finally, the bot sleeps for a random interval until the next action. 
To avoid behaviors typical of spam bots that violate Twitter's polices, the follow and unfollow actions have additional constraints regarding the ratio between friends and followers. The constraints are mutually exclusive, so that if one of these two actions fails due to a constraint not being satisfied, the other action is performed.
Details about actions, sources, their associated probabilities, and constraints can be found in Supplementary Methods.

The only difference among the \drifters{} was the way their friend lists were initialized. This was our experiment's independent variable, or treatment. 
We started from Twitter accounts associated with established and active news sources with different political alignments. 
While mapping the political spectrum to a one-dimensional scale is reductive, we obtain manageable experimental treatments by selecting five sources: 
\textit{The Nation} (left), \textit{The Washington Post} (center-left), \textit{USA Today} (center), \textit{The Wall Street Journal} (center-right), and \textit{Breitbart News} (right). 
These sources were selected because they are among the most popular on Twitter, as well as among the most followed by users aligned with different portions of the U.S. political spectrum (see Supplementary Table~1). The choice of the five accounts, however, was not based on maximizing any single criterion such as popularity. 
The 15 \drifters{} were divided into five groups so that each of three bots in the same group started by following the same source account.
The friend list of each \drifter{} was then expanded by following a random sample of five English-speaking friends of the first friend, and a random sample of five English-speaking followers of the first friend --- 11 accounts in total.

\subsection*{Political alignment metrics and news feed bias}
\label{sec:political_alignment}

Given our goal of gauging political bias, we need to measure the political alignment of tweets and accounts within the liberal-conservative spectrum. 
This alignment is operationally defined by a score between $-1$ (liberal) and $+1$ (conservative).
We adopt two independent approaches, one based on hashtags and one on links, to ensure the robustness of our results.
Both approaches start with assigning political alignment scores to entities that may be present in tweets, namely hashtags and links. 
See Supplementary Methods for details about how these entities are extracted from tweets. 
The entity scores are then averaged at the tweet level to obtain alignment scores for the tweets, and further at the user level to measure the political alignment of users.

The hashtag-based approach relies on hashtags (keywords preceded by the hash mark \texttt{\#}) commonly included by users in their social media posts because they are concise and efficient ways to label topics, ideas, or memes so that others can find the messages. Hashtags are often used to signal political identities, beliefs, campaigns, or alignment~\cite{cota2019quantifying}.
We apply the word2vec algorithm~\cite{mikolov2013distributed} to assign political alignment scores to hashtags in a semi-supervised fashion.
Word2vec maps words in text to continuous vector representations, which have been shown to capture semantic relations between the words~\cite{kozlowski2019geometry}.
The axis between a pair of carefully selected word vectors can encode a meaningful cultural dimension and an arbitrary word's position along this axis reflects its association with this cultural dimension. 
Using hashtags as words, we look for an axis representing the political alignment in the embedding vector space.
We leverage a dataset of political tweets collected during the 2018 U.S. midterm elections~\cite{yang2019bot}.
The hashtags from the same tweet are grouped together as a single sentence and fed to the word2vec algorithm to obtain vector representations for the hashtags.
We remove hashtags appearing less than five times in the dataset, which may be rare misspelling or too uncommon to obtain a reliable signal. Note that common variations and misspellings of a hashtag are treated similarly to the original one. 
After filtering, we end up with 54,533 hashtag vectors.
To define the political alignment axis, we choose \texttt{\#voteblue} and \texttt{\#votered} as two poles because they show clear alignment with U.S. liberal and conservative political orientations, respectively.
The rest of the hashtags are then projected onto this axis, and the relative positions, scaled into the interval $[-1, 1]$, are used to measure the political alignment where negative/positive scores indicate left/right alignment.

The link-based approach considers links (URLs) commonly used to share news and other websites in  social media posts, for the purpose of spreading information, expressing opinions, or flagging identity, especially around political matters.
Many websites show clear political bias.
If the political ideology of a set of social media users is known, one can use it in conjunction with knowledge of the websites they share and like on the platform to infer the political bias of the sources~\cite{Bakshy2015}. 
Therefore the websites (domains) extracted from links provide us with another convenient proxy for the political alignment of tweets and users. 
To assess the political alignment of a website, we start with a dataset of 19 thousand ideologically diverse news sources, where each domain is assigned a score reflecting its political alignment in the liberal-conservative range $[-1, +1]$. These scores are obtained from the sharing activity of Twitter accounts associated with registered U.S. voters~\cite{robertson2018auditing,Robertson2018dataset}. 
For each link found in the tweets, we matched the domain to the list to obtain a score. See Supplementary Methods for additional details.

We further aggregate the political alignment scores of the tweets, obtained using either hashtags or links, at the account level.
We examine different types of political alignment for accounts, each measured on a daily basis.
The political alignment to which a \drifter{} is exposed, $s_h$, is computed by averaging the scores of 50 most recent tweets from its home timeline.
We also evaluate the political alignment expressed by an account by averaging recent tweets they post. We measure this expressed alignment $s_u$ for each \drifter{} using its most recent 20 tweets. 
We use $s_f$ to represent the  political alignment expressed by the friends of each \drifter{}, using their 500 collective tweets. 
In Supplementary Methods we detail how political alignment scores are calibrated so that a value of zero can be interpreted as aligned with the political center. 

Since $s_f$ represents the political alignment expressed by the friends of a \drifter{} and $s_h$ represents the alignment of the posts to which the \drifter{} is  exposed in its home timeline, the difference $s_h - s_f$ can be used to measure potential political bias in Twitter's news feed.

\subsection*{Identification of low-credibility content}
\label{sec:low_credibility}

In evaluating the credibility of the content to which \drifters{} are exposed, we focus on the sources of shared links to circumvent the challenge of assessing the accuracy of individual news articles~\cite{Lazer-fake-news-2018}.
Annotating content credibility at the domain (website) level rather than the link level is an established approach in the literature~\cite{shao2018spread,grinberg2019fake,guess2019less,pennycook2019fighting,bovet2019influence}.

A low-credibility source is one that exhibits extreme bias, propaganda, conspiratorial content, or fabricated news. 
We use a list of low-credibility sources compiled from several  recent research papers. 
Specifically, we consider a source as low-credibility if it fulfills any one of the following criteria: (1)~labeled as low-credibility by Shao \textit{et al.}~\cite{shao2018spread}; (2)~labeled as ``Black,'' ``Red,'' or ``Satire'' by Grinberg \textit{et al.}~\cite{grinberg2019fake}; (3)~labeled as ``fake news'' or ``hyperpartisan'' by Pennycook \textit{et al.}~\cite{pennycook2019fighting}; or (4)~labeled as ``extreme left,'' ``extreme right,'' or ``fake news'' by Bovet \textit{et al.}~\cite{bovet2019influence}. 
This provides us with a list of 570 sources. 

To measure the percentage of low-credibility links, we extracted the links from the home timelines of the \drifters{} (expanding those that are shortened) and then cross-referenced them with the list of low-credibility sources.

\subsection*{Echo chambers}
\label{sec:echo_chamber}

We wish to measure the density and transitivity of each \drifter{} bot's ego network. Since reconstructing the full network of friends and followers of each bot is prohibitively time consuming due to the Twitter API's rate limits, we approximated each bot's ego network by sampling 100 random neighbors from a list of the latest 200 friends and 200 followers returned by the Twitter API. We then checked each pair of sampled neighbors for friendship. We added an undirected edge if there was a follower/friend link in either direction, so that the sampled ego network is undirected and unweighted. Finally, we computed the density and transitivity of each ego network~\cite{journals/jgaa/SchankW05}. 

Since transitivity is correlated with density, we also normalized the transitivity by the average transitivity of 30 shuffled networks generated by a configuration model that preserves the degree sequence of the original ego network.
We replace any self-loop and parallel edges, generated by the configuration model, by random edges. 

\subsection*{Data availability}

To ensure reproducibility of the experiment in this study, we share our code that generates a database of Twitter content. We do not share the resulting raw data to comply with Twitter terms that prohibit sharing content obtained from the Twitter API with third parties. We do provide code to process the raw data into an intermediate data format that includes all the derived information necessary for analysis. User and tweet IDs are anonymized to protect subject privacy. This allows other researchers to reproduce our results and/or compare their results. 

The data are available in a public repository at Github (\url{github.com/IUNetSci/DrifterBot}) and at Zenodo (\url{doi.org/10.5281/zenodo.4750190}) \cite{Drifter2021dataset}. 
Source data are also provided with this paper.

\subsection*{Code availability}

All of the code used to run the experiment and produce the figures in this manuscript is available in a public repository at Github (\url{github.com/IUNetSci/DrifterBot}) and at Zenodo (\url{doi.org/10.5281/zenodo.4750190}) \cite{Drifter2021dataset}.
The repository lists dependencies on external libraries, such as twurl, chatterbot, gensim, and the Botometer Pro API and Python client library.

\bibliographystyle{unsrt}  
\bibliography{main_paper}  

\newpage

\subsection*{Acknowledgements}

We are grateful to Paul Cheung for a conversation that inspired the \drifter{} experiment, and to Eni Mustafaraj for helpful suggestions. This work was supported in part by Knight Foundation and Craig Newmark Philanthropies. Any opinions, findings, and conclusions or recommendations expressed in this material are those of the authors and do not necessarily reflect the views of the funding agencies.

\subsection*{Author contributions}

F.M. and W.C. designed the study. W.C. and D.P. implemented the bot activities and data collection. K.C.Y. developed the hashtag embedding method to infer political alignment scores. All authors analyzed the data and wrote the report. 

\subsection*{Competing interests}

The authors declare no competing interests.

\newpage

\begin{figure}[!ht]
    \centering
    \includegraphics[width=0.7\columnwidth]{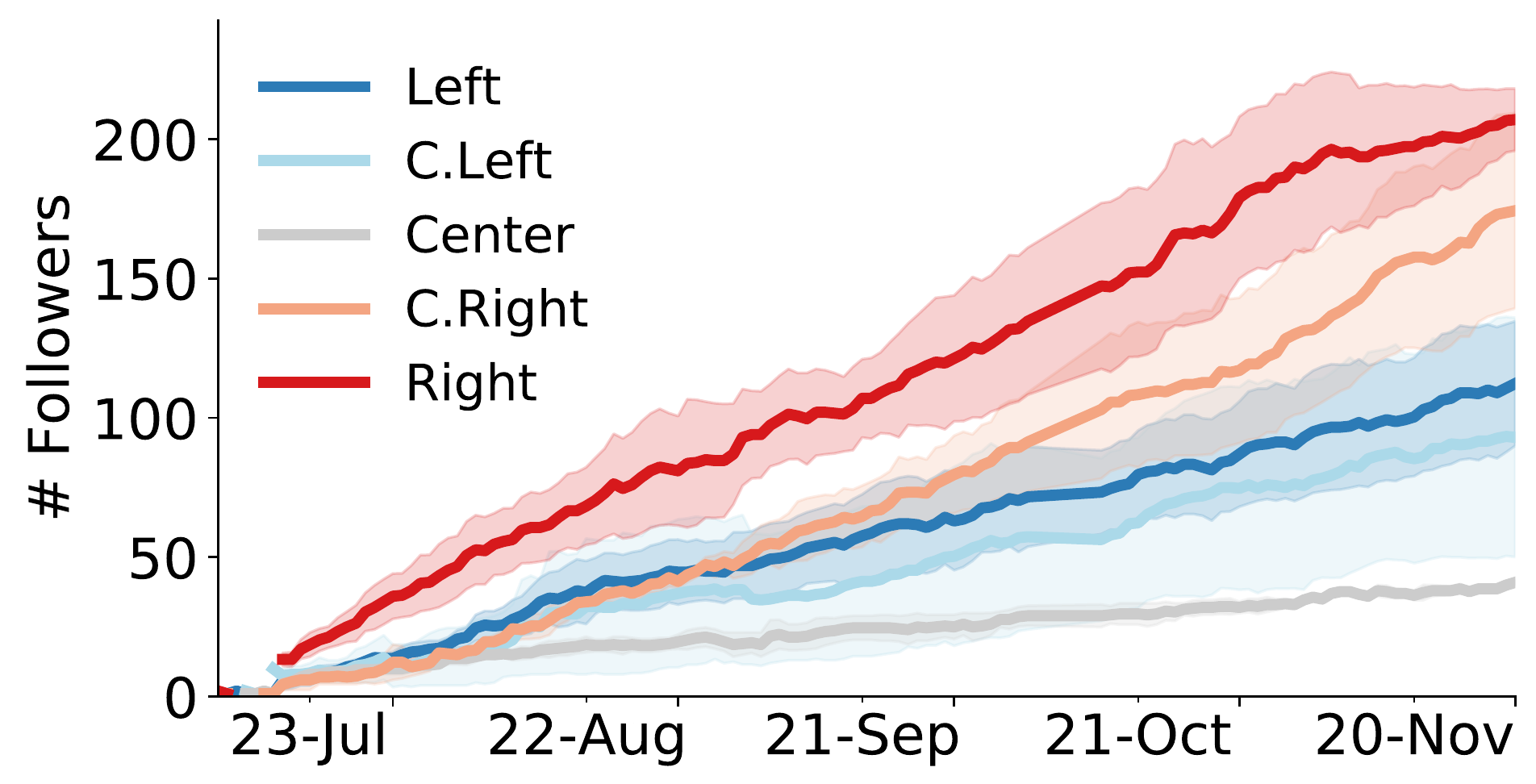}
    \caption{\textbf{Growth in followers.} The x axis displays the duration of the experiment in 2019 and the y axis reports the average numbers of followers of different \drifter{} groups. Colored confidence intervals indicate $\pm 1$ standard error. Source data are provided as a Source Data file.}
    \label{fig:followers}
\end{figure}

\begin{figure}[!ht]
    \centering
    \includegraphics[width=.9\columnwidth]{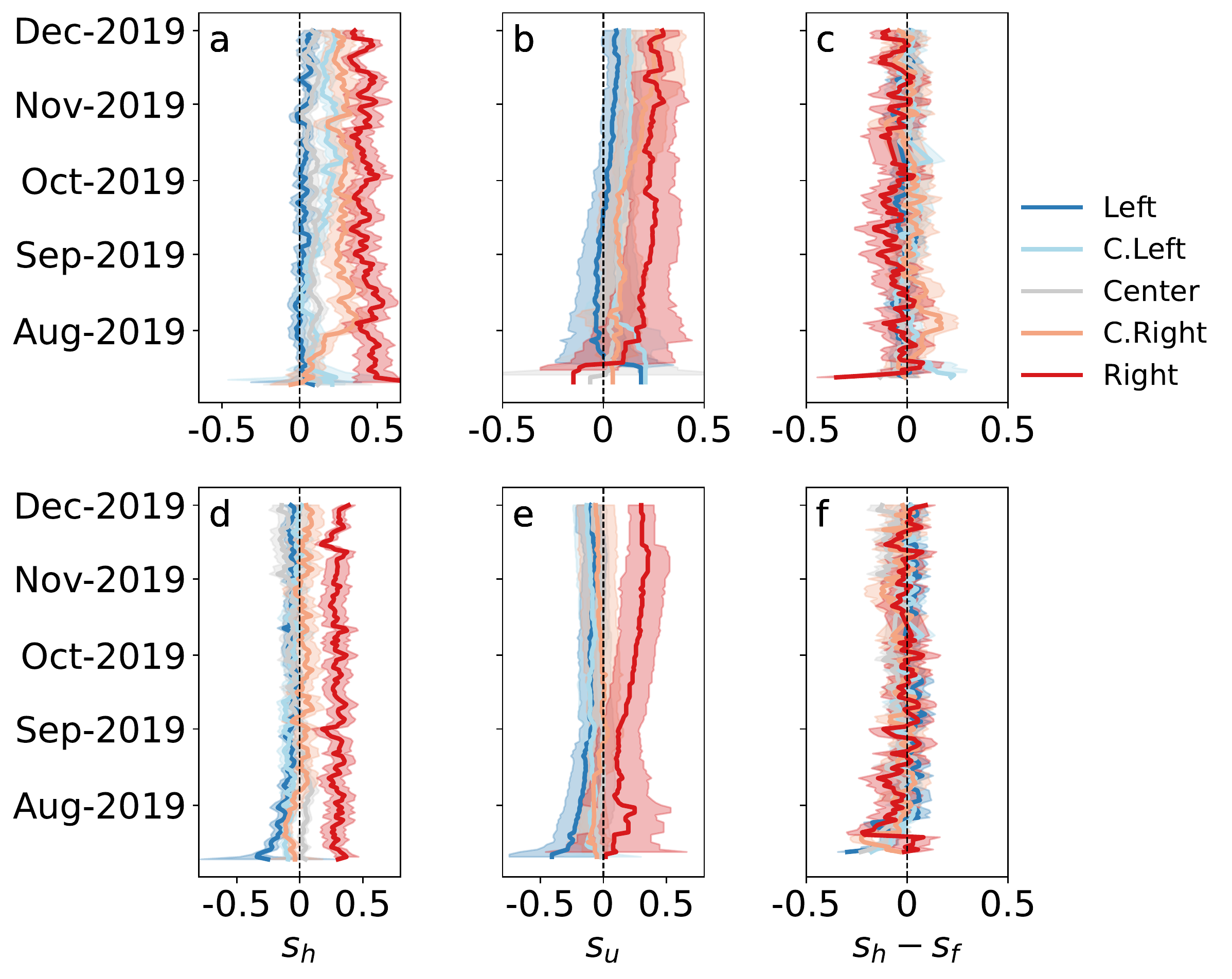}
    \caption{
    \textbf{Time series of political alignment.}  
    Negative alignment scores mean left-leaning and positive scores mean right-leaning. 
    Alignment scores $s_h$ are calculated from content to which \drifters{} are exposed in their \textit{home} timelines, based on hashtags (\textbf{a}) and links (\textbf{d}). 
    Alignment scores $s_u$ are calculated from content expressed by \drifter{} posts in their \textit{user} timelines, based on hashtags (\textbf{b}) and links (\textbf{e}). 
    The difference $s_h - s_f$ measures news feed bias experienced by \drifters{}, where the political alignment score $s_f$ is derived from the content generated by \textit{friends}, based on hashtags (\textbf{c}) and links (\textbf{f}).
    Missing values are replaced by preceding available ones. Colored confidence intervals indicate $\pm 1$ standard error.
    Source data are provided as a Source Data file.
    Plots for each individual \drifter{} are in Supplementary Figures 2--4.  
    }
    \label{fig:timeline_valence}
\end{figure}

\clearpage

\appendix

\setcounter{page}{1}
\renewcommand{\figurename}{Supplementary Figure}
\renewcommand{\tablename}{Supplementary Table}
\setcounter{figure}{0}   
\setcounter{table}{0} 

\section*{Neutral bots probe political bias on social media}

Wen Chen, Diogo Pacheco, Kai-Cheng Yang, and Filippo Menczer

\section*{Supplementary information}

\subsection*{Supplementary methods}

\subsubsection*{Seed accounts}

While initializing the friend lists of the drifters, we aim to select five Twitter accounts that are associated with established, active, and popular U.S. news sources and that span the full range of the U.S. political spectrum.
Supplementary Table~\ref{table:media_twitter_stats} lists the ten news sources with most followers on Twitter in each political leaning bin.
The political leaning of the sources in the table is obtained from the AllSides media bias rating list (\url{www.allsides.com/media-bias/media-bias-ratings}). 
Note that \textit{The Wall Street Journal} is categorized as both center and center-right by the list.
We assign it to the center-right category.
The selected accounts (in bold) are among the most popular news sources on Twitter.

We also wish to verify that the seed accounts are popular among active Twitter users who are  politically aligned with those sources. To this end, we started with a 10\% random sample of public tweets on August 1, 2019, comprising about 36M tweets from 14M unique accounts. We sampled 500k of these accounts; for each of them, we calculated the bot score using the BotometerLite tool.\footnote{Yang, K. C., Varol, O., Hui, P. M., \& Menczer, F. (2020). Scalable and generalizable social bot detection through data selection. In Proceedings of the AAAI Conference on Artificial Intelligence (Vol. 34, No. 01, pp. 1096-1103)} We removed likely bots (those with bot score above 0.5) as well as non-English accounts, leaving 151,570 accounts. 
We extracted tweets by those accounts from the 10\% random sample during one week around August 1, 2019.
We used the links shared in those tweets to assign a political score to each of those accounts (see Methods). We filtered out accounts for which we could not assign a score. That left 26,304 accounts with a political score. 
We grouped these accounts into five political bins using thresholds -1, -0.5, -0.1, +0.1, +0.5, and +1, yielding groups of 712, 5,280, 11,422, 8,237, and 653 accounts in the left, center-left, center, center-right, and right bins, respectively.
By examining the friends of these accounts, we eliminated those who did not follow any of the news sources in the AllSides list. 4,884 accounts remained: 187, 1,362, 1,623, 1,444, and 268 in the five groups, respectively.
Finally, Supplementary Table~\ref{table:media_twitter_stats} reports the proportions of accounts in each of these groups who follow the top sources in the AllSides list.
This confirms that the seed sources are among the most followed by active Twitter users who are politically aligned with those sources. 

\begin{table}
\caption{
\textbf{News sources with most followers on Twitter for each political bin.} Data collected on October 22, 2020. We also consider a sample of active accounts in each group and report the percentage who follow each source. Seed accounts used for initializing the friend lists of the drifters are highlighted in bold.
}
\centering
\resizebox{\columnwidth}{!}{
\footnotesize
\begin{tabular}{lllrr}
\hline
Leaning & News source & Twitter handle &  \multicolumn{2}{c}{Followers}\\
 &  &  &  Total & Within group\\
\hhline{=====}
\multirow{10}{*}{Left}     & HuffPost                         & \mention{HuffPost}        &  11.4M   & 34\%\\
     & The New Yorker                   & \mention{NewYorker}       &   8.9M   & 30\%\\
     & MSNBC                            & \mention{MSNBC}           &   3.6M  & 29\%\\
     & Mother Jones                     & \mention{MotherJones}     & 850.7k  & 29\%\\
     & Vox                              & \mention{voxdotcom}       & 983.6k  & 21\%\\
     & Slate                            & \mention{Slate}           &   1.8M  & 20\%\\
     & \textbf{The Nation}              & \mention{thenation}       &   1.2M  & 20\%\\
     & The Daily Beast                  & \mention{thedailybeast}   &   1.3M  & 19\%\\
     & Newsweek                         & \mention{Newsweek}        &   3.4M  & 14\%\\
     & BuzzFeed News                    & \mention{BuzzFeedNews}    &   1.3M  & 13\%\\
\hline
\multirow{10}{*}{C.~Left}  & The New York Times               & \mention{nytimes}         &  47.5M & 48\%\\
  & \textbf{The Washington Post}     & \mention{washingtonpost}  &  16.4M & 43\%\\
  & CNN News                         & \mention{CNN}             &  50.2M & 38\%\\		
  & The Guardian                     & \mention{guardian}        &   9.1M & 33\%\\
  & The Economist                    & \mention{TheEconomist}    &  25.0M & 29\%\\
  & TIME                             & \mention{TIME}            &  17.4M & 28\%\\
  & ABC News                         & \mention{ABC}             &  15.9M & 23\%\\
  & NBC News                         & \mention{NBCNews}         &   7.8M  & 23\%\\
  & CBS News                         & \mention{CBSNews}         &   7.7M & 20\%\\
  & Bloomberg                        & \mention{business}        &   6.7M & 17\%\\
\hline
\multirow{9}{*}{Center}   & Associated Press                 & \mention{AP}              &  14.4M & 34\%\\
   & BBC News                         & \mention{BBCWorld}        &  29.3M & 31\%\\
   & Reuters                          & \mention{Reuters}         &  22.4M & 28\%\\
   & NPR News                         & \mention{NPR}             &   8.4M & 24\%\\
   & The Hill                         & \mention{thehill}         &   3.9M & 15\%\\
   & \textbf{USA Today}               & \mention{USATODAY}        &   4.1M & 14\%\\
   & Axios                            & \mention{axios}           & 494.4k & 6\%\\
   & Real Clear Politics              & \mention{RealClearNews}   & 183.5k & 3\%\\
   & Christian Science Monitor    & \mention{csmonitor}       &  79.1k     & 2\%\\
\hline
\multirow{10}{*}{C.~Right} & Fox News                         & \mention{FoxNews}         &  19.8M & 30\%\\
 & \textbf{The Wall Street Journal} & \mention{WSJ}             &  18.1M & 22\%\\
 & New York Post                    & \mention{nypost}          &   1.9M & 18\%\\
 & The Epoch Times                  & \mention{EpochTimes}      & 319.0k & 17\%\\
 & NewsMax                         & \mention{newsmax}         & 223.1k & 11\%\\
 & Washington Examiner              & \mention{dcexaminer}      & 272.9k & 8\%\\
 & Market Watch                     & \mention{MarketWatch}     &   3.8M & 5\%\\
 & The Washington Times             & \mention{WashTimes}       & 406.4k & 5\%\\
 & Reason                           & \mention{reason}          & 260.2k & 3\%\\
 & The American Conservative        & \mention{amconmag}        &  53.0k & 1\%\\
\hline
\multirow{10}{*}{Right}    & One America News                 & \mention{OANN}            &   1.1M & 68\%\\
    & \textbf{Breitbart News}          & \mention{BreitbartNews}   &   1.5M & 61\%\\
    & The Daily Caller                 & \mention{DailyCaller}     & 775.2k & 44\%\\
    & The Blaze                        & \mention{theblaze}        & 762.2k & 31\%\\
    & The Federalist                   & \mention{FDRLST}          & 277.0k & 29\%\\
    & The Daily Wire                   & \mention{realDailyWire}   & 493.9k & 23\%\\
    & National Review                  & \mention{NRO}             & 350.5k & 20\%\\
    & Fox News Opinion                 & \mention{FoxNewsOpinion}  & 155.8k & 10\%\\
    & CBN News                         & \mention{CBNNews}         & 148.9k & 9\%\\
    & Daily Mail US                    & \mention{DailyMail}       & 309.4k & 4\%\\
\hline
\end{tabular}
}
\label{table:media_twitter_stats}
\end{table}

\subsubsection*{\Drifter{} actions and probabilities}

An \textit{action} is performed upon a sentence, an existing tweet, or a user. These inputs are selected from \textit{sources} that are described below. A \drifter{} can perform the following actions:

\begin{itemize}
    \item \textbf{Tweet} -- post a sentence from \textit{Random Quotes}, \textit{Trends}, or \textit{Home Timeline}. For \textit{Trends}, the sentence is the text of the selected tweet. For \textit{Home Timeline}, the sentence is obtained by concatenating a short phrase from a manually compiled list (e.g., ``Wow!'' or ``Maybe so.'') with the link of the selected tweet. This emulates a quoted tweet. 
    
    \item \textbf{Retweet} -- select a tweet from \textit{Trends}, \textit{Home Timeline}, or a list of \textit{Tweets Liked by Friends}, and retweet it.
    
    \item \textbf{Like} -- like a tweet selected in the same way as for a \textit{Retweet}.
    
    \item \textbf{Reply} -- reply to a tweet from the \textit{Mention Timeline}. The reply is generated using the ChatterBot library (\url{chatterbot.readthedocs.io}). In case of failure, the reply is a random phrase from the precompiled list described above.
    
    \item \textbf{Follow} -- select a user to follow from the list of \textit{Followers}, \textit{Friends of Friends}, users who posted \textit{Tweets liked by Friends}, or users who posted tweets in the \textit{Home Timeline}.
    
    \item \textbf{Unfollow} -- select a user to unfollow from the latest 200 in the list of \textit{Friends}. 
\end{itemize}
 
Input elements for actions are selected from candidate lists that we call \textit{sources}. The selection is random with uniform probability distribution unless otherwise explained below. 
Due to limitations of the Twitter APIs, we imitate some basic mechanisms offered by the platform, such as suggestions to follow friends of friends.
Sources are defined as follows:

\begin{itemize}
    \item \textbf{Random Quotes} -- sentences obtained from a random quote API (\url{api.quotable.io/random}).
    
    \item \textbf{Mention Timeline} -- the latest 10 tweets in the mention timeline. If the \drifter{} replied to any mentions in the past, this source only considers subsequent tweets.
    
    \item \textbf{Friends of Friends} -- the model randomly selects three friends of the \drifter{} and requests their latest 5,000 friends, ignoring those that are already friends of the \drifter{}. A new friend is selected from the combined list with probability proportional to the occurrences in the list, to favor friends of multiple friends.
    
    \item \textbf{Friends} -- most recent 200 friends. The user is selected from this list at random, but older friends are more likely to be unfollowed. We implement this mechanism by ranking friends chronologically; the latest friend has rank one. The unfollow probability is proportional to the rank. The initial friend can never be unfollowed.
    
    \item \textbf{Followers} -- most recent 200 followers. 
    
    \item \textbf{Trends} -- list obtained by randomly selecting three trending topics in the U.S. and fetching the top five tweets in each topic by the default ranking.
    
    \item \textbf{Tweets Liked by Friends} -- start from the latest 15 tweets from the home timeline. Select a random subset of at most ten friends who posted these tweets. Select the three latest tweets liked by each of the selected friends, excluding any by the \drifter{} itself. Select one tweet at random from this combined list. Depending on the selected action, the source can return the tweet itself (for \textbf{Retweet} or \textbf{Like}) or its author (for \textbf{Follow}). 
    
    \item \textbf{Home Timeline} -- the latest 15 tweets in the home timeline.
\end{itemize}

\begin{table}[b!]
\centering
\caption{\textbf{Probabilities of actions and sources in the \drifter{} bot behavior model.} The probabilities of the actions add up to one, and so do the conditional probabilities of the sources given each action.}
\begin{tabular}{lc|lc}
\hline
Action & $P(\text{Action})$ & Source & $P(\text{Source} | \text{Action})$ \\
\hhline{====}
Reply & 0.05 & Mention Timeline & 1.0 \\
\hline
\multirow{3}{*}{Tweet}  & \multirow{3}{*}{0.15}  & Random Quotes & 0.3 \\
   & & Trends & 0.3 \\
   & & Home Timeline & 0.4\\
\hline
\begin{tabular}{@{}l@{}}Retweet \\  Like \end{tabular} &
\begin{tabular}{@{}l@{}}0.1 \\  0.35 \end{tabular}& \begin{tabular}{@{}l@{}}Trends \\ Home Timeline \\ Tweets Liked by Friends\end{tabular} & 
\begin{tabular}{@{}l@{}}0.1 \\ 0.6 \\ 0.3\end{tabular}
\\
\hline
\multirow{4}{*}{Follow} & \multirow{4}{*}{0.25}  & Home Timeline  & 0.2      \\
  & & Tweets Liked by Friends & 0.2 \\
  & & Friends of Friends  & 0.5  \\
  & & Followers & 0.1 \\
 \hline
  Unfollow & 0.1 & Friends & 1.0 \\ \hline
\end{tabular}
\label{tbl:bot_prob}
\end{table}

We list the probabilities used in the bot behavior model in Supplementary Table~\ref{tbl:bot_prob}.
The numbers are inferred from a random sample of Twitter users. 
If the Follow or Unfollow action is selected, a precondition check is triggered. If the Follow precondition is not met, the Unfollow action is performed and viceversa; the two checks cannot both fail. A new friend can only be followed if the number of friends is sufficiently small compared to the number of followers: less than the number of followers plus 113. A friend can only be unfollowed if the \drifter{} has at least 50 friends.

\subsubsection*{Extraction of hashtags and links}

We accessed tweets using the Twitter API. Links (URLs) and hashtags were extracted from \texttt{entities} metadata. Tweets longer than 140 characters are truncated; in these cases, we extracted links and hashtags from the \texttt{extended\_entities} metadata except for 4\% of the tweets, for which this  retrieval process failed.  

Many links are compressed using URL-shortening services. We expanded shortened links via HTTP HEAD requests using a heuristics based on the length of the URL (20 characters of less), allowing multiple redirects with a 10-second timeout.

\subsubsection*{Calibration of political alignment scores}
\label{sec:score_calibration}

We calibrated alignment scores so that positive scores mean right-leaning hashtags/links and negative scores mean left-leaning hashtags/links. 
To this end, we selected the news source account \mention{USATODAY} to have a zero alignment score.
We used the 200 most recent tweets by \mention{USATODAY} in early June to calculate the raw center alignment score $s_c$.
We obtained $s_c = 0.058$ and $s_c = -0.246$ for the link-based and hashtag-based approach, respectively.
The political alignment scores are then calibrated by
$s = \frac{1}{N} \sum^{N}_{i} (t_i - s_c)$,
where $t_i$ is the score for tweet $i$ and $N$ is the number of tweets across which the score is aggregated.

\subsubsection*{Statistical analyses}

All t-tests in our analyses are two-sided. In cases where we compare five groups of \drifters{}, a Bonferroni correction for multiple comparisons can be applied by dividing the significance level by $\binom{5}{2}=10$. The main results in the paper are significant at the 0.05 level (0.005 after the correction).

\subsection*{Supplementary notes}

\subsubsection*{Comparisons of follower growth rates and confounding factors}

There are multiple ways to test whether \drifters{} in one group gain followers significantly faster than those in another group (Fig.~\ref{fig:followers}). 
The method reported in the main text focuses on the daily follower growth for each \drifter{} bot. We record the follower count on a daily basis in our experiment, with a few exceptions due to technical issues. We calculate the daily growth rate for any two consecutive observations of the follower count. We then combine the data points from each group and use t-tests to compare different groups ($n$ between 373 and 389).

Here we report on analyses based on two additional methods. In the first, we first combine the raw observations (follower-date pairs) from the \drifters{} within the same group and then combine them across two groups to be compared, using a dummy variable to distinguish them. Finally we apply linear regression to this combined data set with an extra interaction term between elapsed time and the dummy variable. The coefficient of the interaction term indicates whether the growth rates between the two groups are significantly different ($n=782$, $p<0.001$ comparing Left vs. Center and $n=779$, $p<0.001$ comparing Right vs. Left).

In the last method, we use linear regression to estimate the follower growth rate for each \drifter{}. We then use a two-sided t-test to compare the estimated growth rates of two different groups ($d.f.=4, t=5.43, p=0.006$ comparing Right vs. Center; $d.f.=4, t=2.71, p=0.054$ comparing Left vs. Center; and $d.f.=4, t=2.60, p=0.060$ comparing Right vs. Left). 
All three methods yield consistent results, which are more significant for the two methods with more degrees of freedom.

\begin{figure}
    \centering
    \includegraphics[width=0.9\columnwidth]{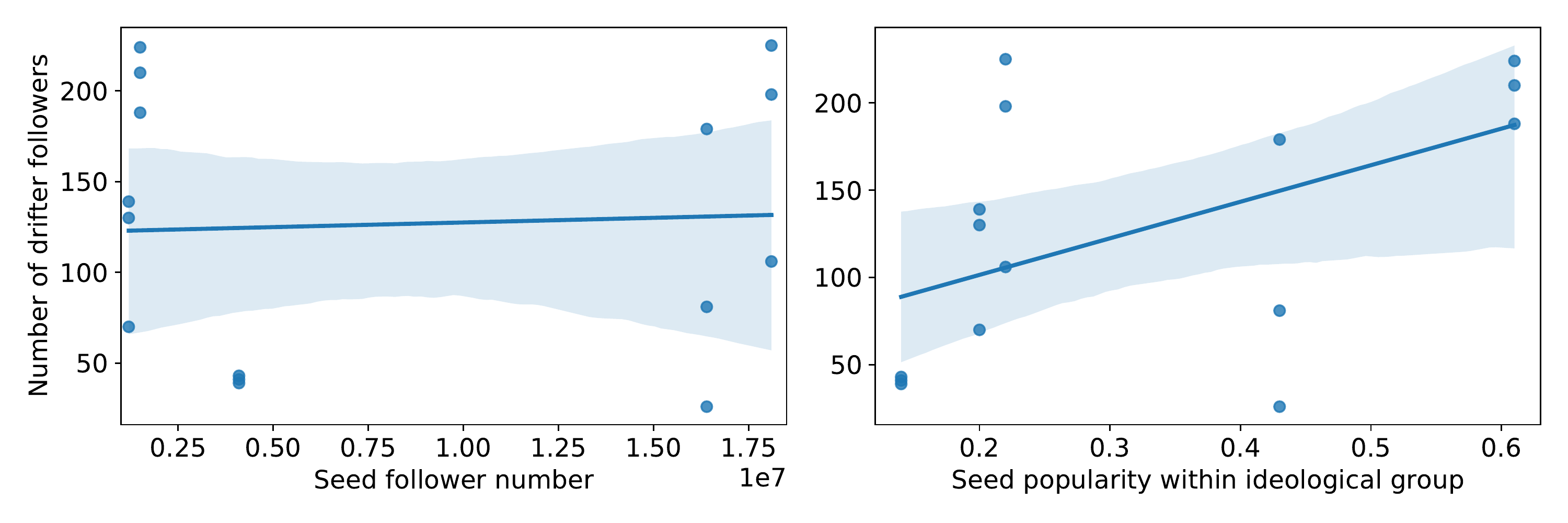}
    \caption{\textbf{Confounding factors for \drifter{} influence.} The scatter plots show the relationships between the number of followers of each \drifter{} and (left) the number of followers and (right) the within-group popularity of its first friend. The number of followers of the \drifters{} was measured on December 2, 2019. The seed popularity measures are taken from Supplementary Table~\ref{table:media_twitter_stats}; the methods to calculate overall and within-group popularity are documented in supplementary methods above. Shaded areas highlight the 95\% confidence intervals around least-squared linear fits (solid lines). Source data are provided as a Source Data file.}
    \label{fig:confounding}
\end{figure}

The differences in influence among \drifters{} could be affected by the popularity of the seed accounts. Supplementary Fig.~\ref{fig:confounding} shows the correlation between \drifter{} influence and two measures of popularity of their respective seed accounts. We find no significant correlation between the numbers of followers of \drifters{} and seed accounts (Pearson's $r=0.05$, $p=0.850$). However, the \drifter{} influence is correlated with the popularity of the seeds among active accounts with similar political alignment (Pearson's $r=0.52$, $p=0.049$).

\subsubsection*{Individual political trajectories}

Starting with the political alignment estimations, Fig.~\ref{fig:timeline_valence} in the main text shows the aggregated scores.
Next we provide the individual trajectory of the political alignment for each \drifter{}. Full-resolution vector images of the plots are available in the data repository (\url{github.com/IUNetSci/DrifterBot}). 

\begin{figure}
    \centering
    \begin{overpic}[trim = 0mm 7mm 0mm 0mm, clip,width=0.9\columnwidth]{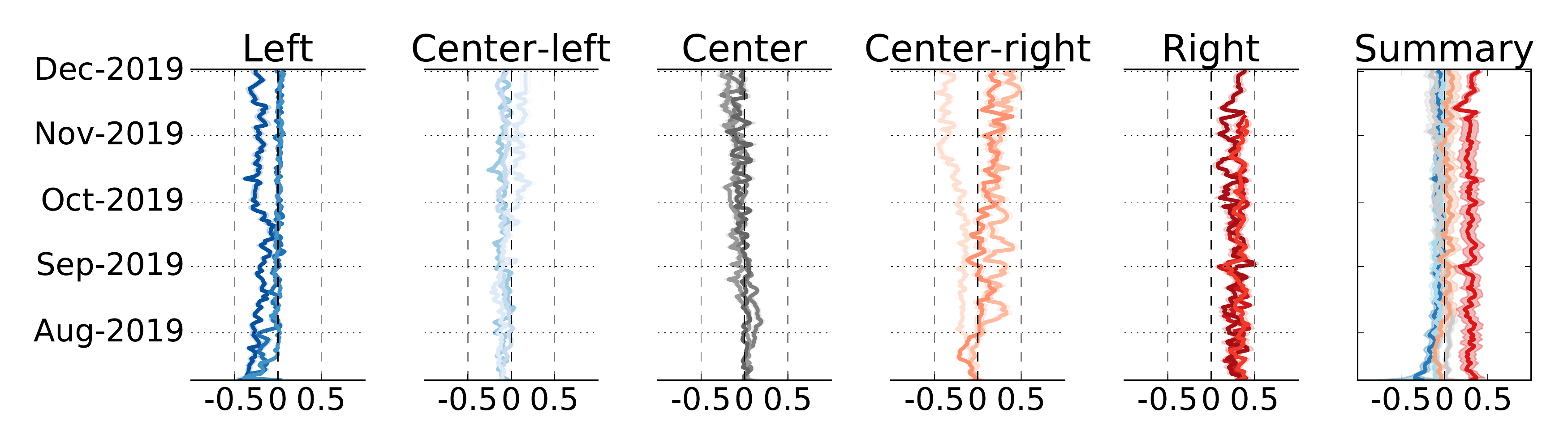}
    \put(-5,12){A}
    \put(-5,-10){B}
    \end{overpic}
    \includegraphics[trim = 0mm 7mm 0mm 7mm, clip,width=0.9\columnwidth]{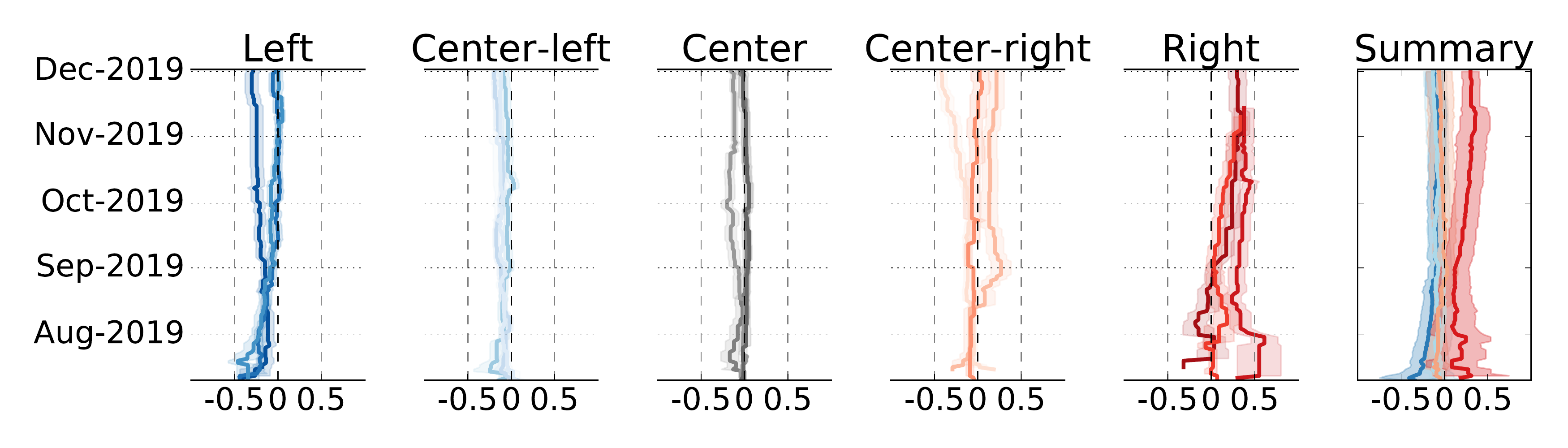}
    \caption{\textbf{Political alignment timelines based on links for all fifteen bots.} A tweet is assigned a score between $-1$ (liberal) and $+1$ (conservative) based on the shared link domains. \textbf{A}~Home timeline: daily average score of the last 50 tweets in the home timeline. \textbf{B}~User timeline: daily average score of the last 20 tweets in the user timeline. The summary represents the average for each group. Colored confidence intervals indicate $\pm 1$ standard error. Source data are provided as a Source Data file.}
    \label{fig:timeline_url_all}
\end{figure}

\begin{figure}
    \centering
    \begin{overpic}[trim = 0mm 7mm 0mm 0mm, clip,width=0.9\columnwidth]{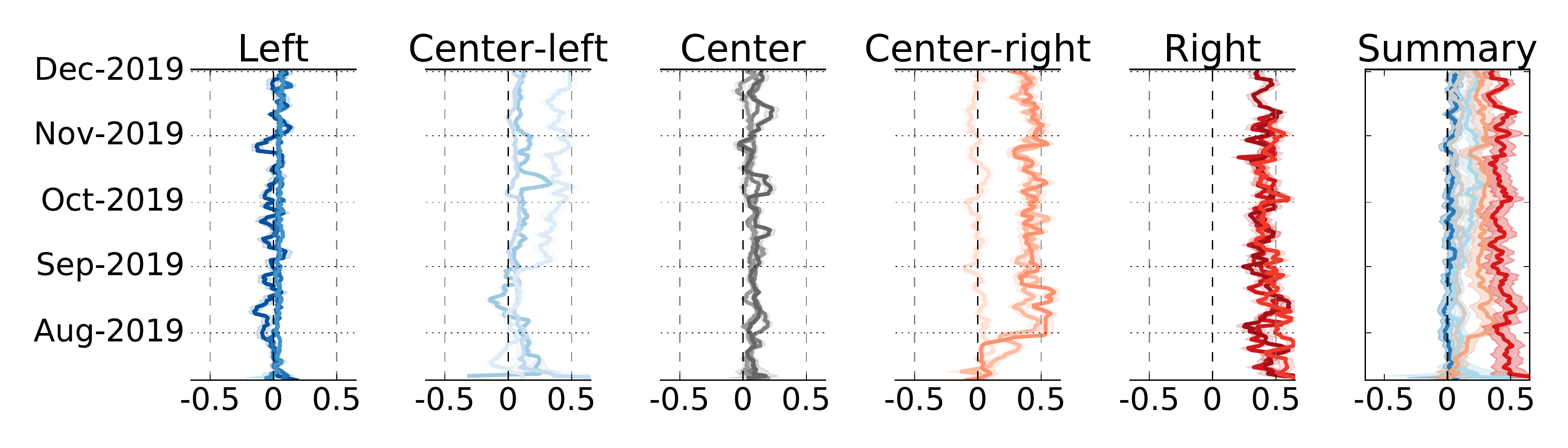}
    \put(-5,12){A}
    \put(-5,-10){B}
    \end{overpic}
    \includegraphics[trim = 0mm 7mm 0mm 7mm, clip,width=0.9\columnwidth]{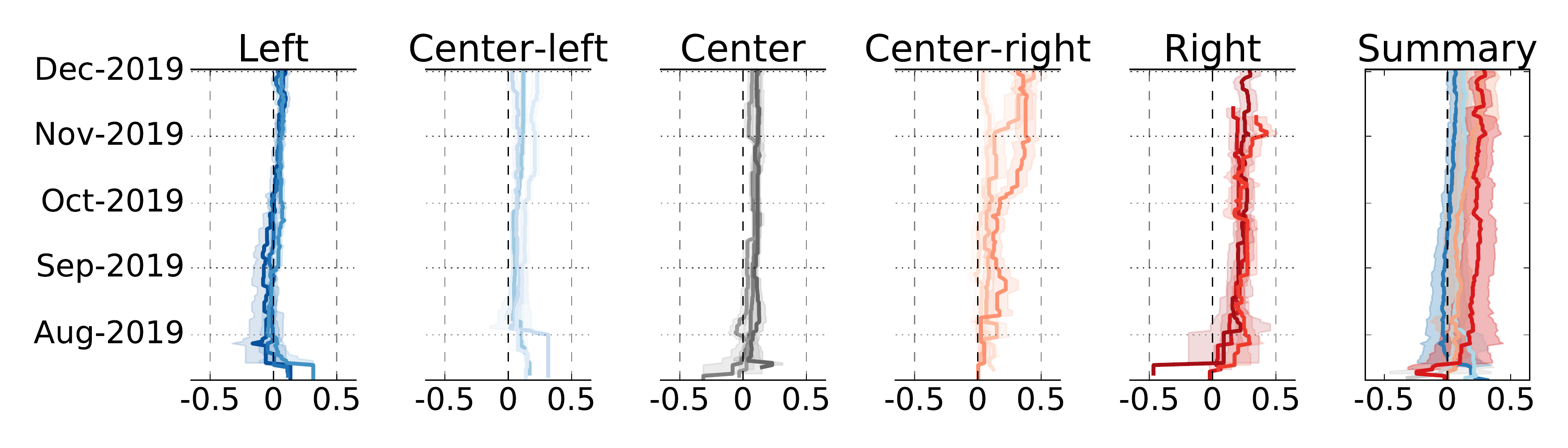}
    \caption{\textbf{Political alignment timelines based on hashtags for all fifteen bots.} A tweet is assigned a score between $-1$ (liberal) and $+1$ (conservative) based on the shared hashtags. \textbf{A}~Home timeline: daily average score of the last 50 tweets in the home timeline. \textbf{B}~User timeline: daily average score of the last 20 tweets in the user timeline. The summary represents the average for each group. Colored confidence intervals indicate $\pm 1$ standard error. Source data are provided as a Source Data file.}
    \label{fig:timeline_hashtag_all}
\end{figure}

\begin{figure}
    \centering
    \begin{overpic}[trim = 0mm 7mm 0mm 0mm, clip,width=0.9\columnwidth]{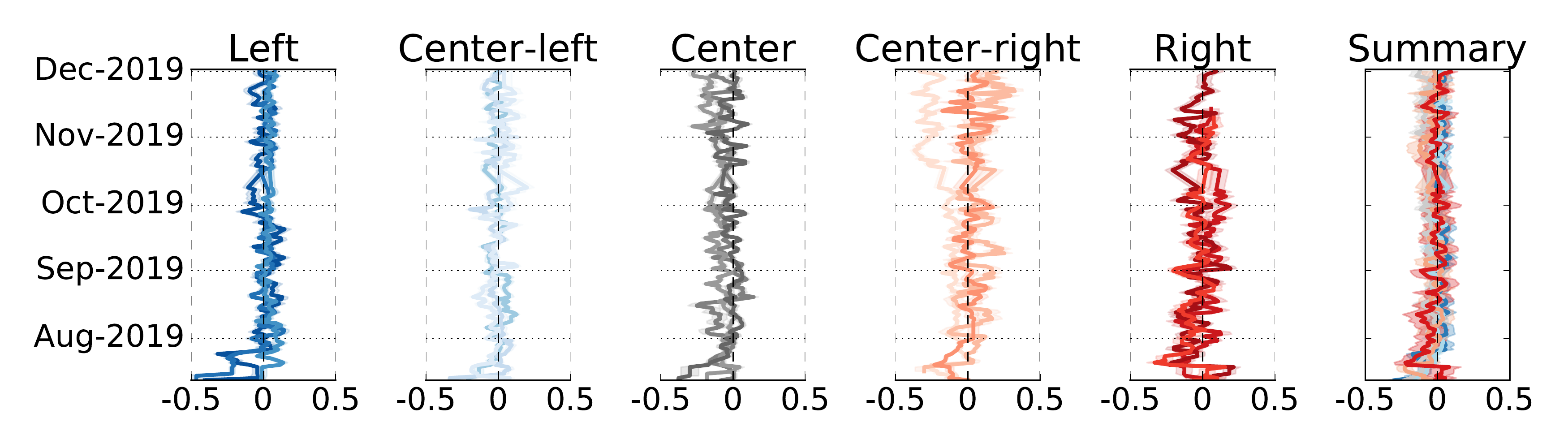}
    \put(-5,12){A}
    \put(-5,-10){B}
    \end{overpic}
    \includegraphics[trim = 0mm 7mm 0mm 7mm, clip,width=0.9\columnwidth]{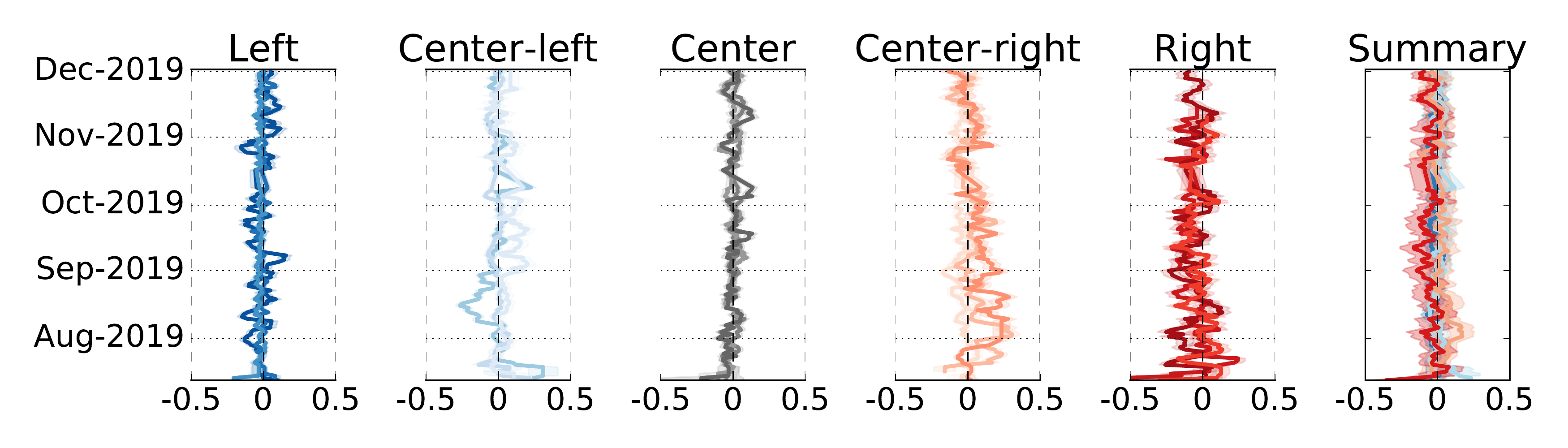}
    \caption{\textbf{News feed bias for all fifteen bots.} Bias is measured by the difference in alignment between the account's home timeline and its friends' user timelines, based on \textbf{A}~links and \textbf{B}~hashtags. The summary represents the average for each group. Colored confidence intervals indicate $\pm 1$ standard error. Source data are provided as a Source Data file.}
    \label{fig:bias_all}
\end{figure}

Supplementary Fig.~\ref{fig:timeline_url_all} shows the results from the link-based approach and  Supplementary Fig.~\ref{fig:timeline_hashtag_all} shows the results from the hashtag-based approach. We observe that the trajectories diverge in several examples, suggesting that the initial conditions do not limit the variability of the evolution. 
Supplementary Fig.~\ref{fig:bias_all} shows the news feed bias computed for each \drifter{} with both methods.
Note that two of the Right \drifters{} were temporarily suspended by Twitter in mid-November 2019, and we neglected to reactivate them until the end of the experiment.

\subsubsection*{Political bias of news feed algorithm}

Supplementary Table~\ref{table:t-test} shows the results of the analysis of bias in the platform's news feed. As discussed in the main text, most effects have small size and are not consistent across link- and hashtag-based methods. 

\subsubsection*{Follow-back rates}

\begin{figure}
    \centering
    \includegraphics[width=0.8\textwidth]{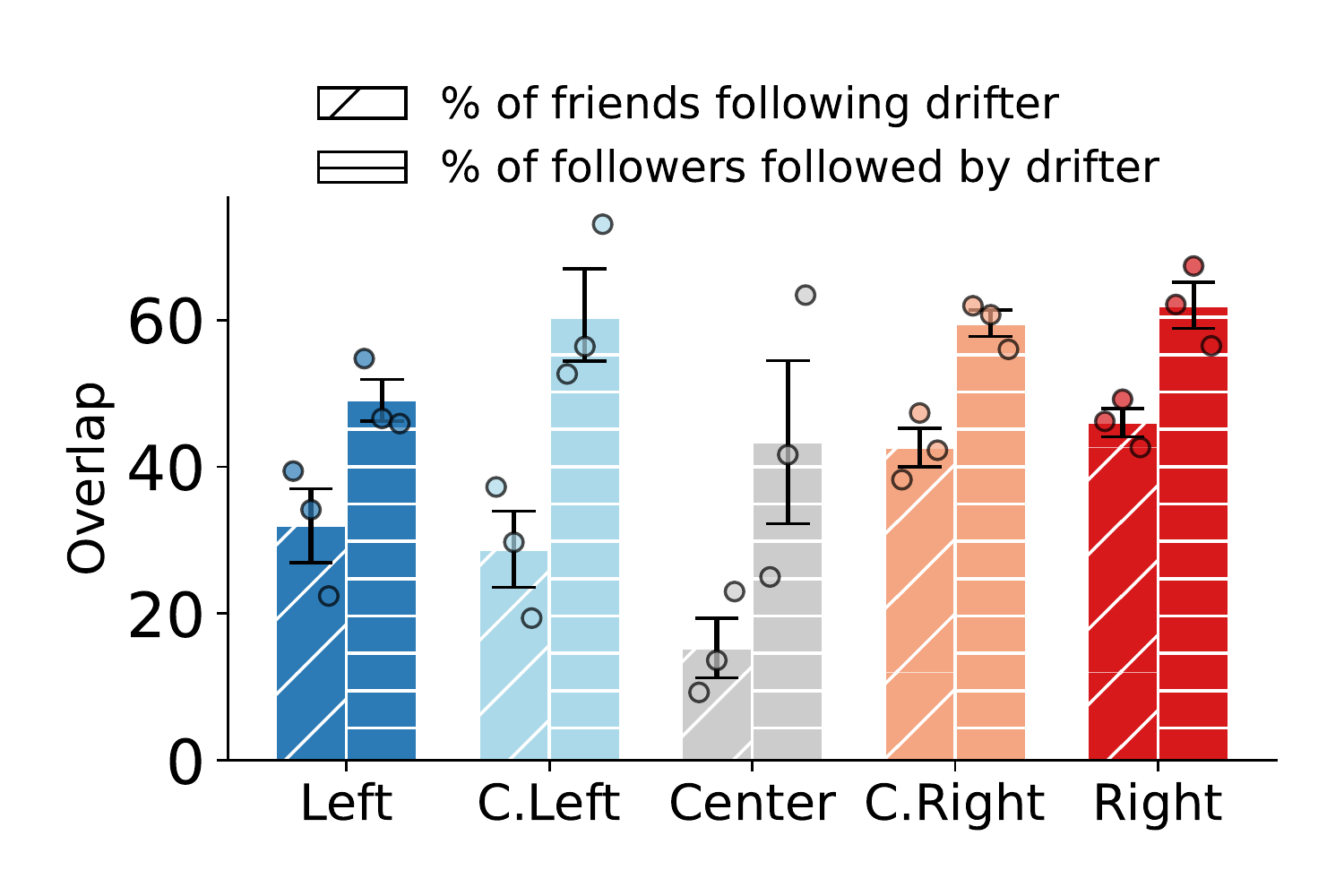}
    \caption{Relative overlap between friends and followers of the \drifters{} in each group. Error bars indicate standard errors ($n=3$ \drifters{} in each group). Source data are provided as a Source Data file.}     
    \label{fig:follow_back}
\end{figure}

Supplementary Fig.~\ref{fig:follow_back} plots the relative overlap between friends and followers of each \drifter{} to examine the reciprocity of the links. We observe a higher follow-back rate for partisan \drifters{}, and especially conservative ones.

\subsubsection*{Descriptive statistics of \drifters{}}

Finally, Supplementary Table~\ref{table:demographic} provides descriptive statistics of the \drifters{}, including the number of friends and followers, number of tweets liked, number of tweets posted (including retweets), number of hashtags and links with alignment scores in posted tweets, and total number of actions taken. 

\begin{table}
\caption{\textbf{Political bias in the platform news feed.} Results of paired two-sided $t$-tests comparing political alignment scores of \drifter{} home timelines and their friends' user timelines. Negative $t$ values indicate left (liberal) bias, positive values indicate right (conservative) bias. Source data are provided as a Source Data file.}
\centering
\begin{tabular}{lccccc}
\hline
Drifters  &   Method & $n$ & $t$ &  $p$       &   Cohen's $d$ \\
\hhline{======}
Left     &  hashtag & 387 & -6.0 &            $<0.001$ &  0.30 \\
C. Left  &  hashtag & 383 &  2.6 &            $0.010$ &  0.13 \\
Center   &  hashtag & 392 &  1.7 &            $0.073$ &  0.09  \\
C. Right &  hashtag & 381 &  4.7 &            $<0.001$ &  0.24 \\
Right    &  hashtag & 352 &-10.6 &            $<0.001$ &  0.56 \\
\hline
Left     &  link    & 391 &   4.1 &            $<0.001$ &  0.21 \\
C. Left  &  link    & 389 &  -2.3 &            $0.021$ &  0.12  \\
Center   &  link    & 393 & -15.2 &            $<0.001$ &  0.76 \\
C. Right &  link    & 385 &  -4.1 &            $<0.001$ &  0.21 \\
Right    &  link    & 352 &  -5.0 &            $<0.001$ &  0.26 \\
\hline
\end{tabular}
\label{table:t-test}
\end{table}

\begin{table}
\caption{\textbf{Descriptive statistics of the \drifters{}.} Data collected until 14 November 2019. Averages and standard deviations are shown for \drifters{} in each group (highlighted) and across groups. Source data are provided as a Source Data file.}
\centering
\setlength{\tabcolsep}{2pt}
\begin{tabular}{lccccccc}
\hline
\Drifters{} & Friends & Followers 
& Tweets &  Likes & Hashtags & Links & Actions \\
\hhline{========}
\multirow{4}{*}{Left} & 159 & 53 & 860 & 1218 & 169&275	& 3413 \\
 & 230 & 118	& 841 &	1179 & 326 & 317 & 3357\\
 & 237 & 124 & 816 & 1177 & 339 & 281 & 3360\\
 & \cellcolor{LightCyan} 209$\pm$35  &	\cellcolor{LightCyan} 98$\pm$32  
 &\cellcolor{LightCyan}	839$\pm$18 &	\cellcolor{LightCyan}1191$\pm$19 
 &\cellcolor{LightCyan}	278$\pm$77 &\cellcolor{LightCyan}	291$\pm$19 
 &	\cellcolor{LightCyan}3377$\pm$26 \\
 \hline
 
 \multirow{4}{*}{C. Left} & 137 &	24&	885	& 1258 & 213 & 300 & 3507\\
 & 269 & 150 & 837 & 1142 & 180 & 251 & 3237\\
 & 184 & 73 & 850 & 1233 & 143 & 257 & 3494\\
 & \cellcolor{LightCyan} 197$\pm$55  & \cellcolor{LightCyan} 82$\pm$52 
 &\cellcolor{LightCyan}	857$\pm$20	  & \cellcolor{LightCyan} 1211$\pm$50	  & \cellcolor{LightCyan} 179$\pm$29	  & \cellcolor{LightCyan} 269$\pm$22	& \cellcolor{LightCyan} 3413$\pm$124\\
 \hline
 
  \multirow{4}{*}{Center} & 151 & 38 & 814 & 1117 & 171 & 232 & 3192\\
 & 	152 & 38 & 827 & 1236 & 182 & 261 & 3446\\
 & 148 & 34 & 908 & 1195 & 158 & 265 & 3581\\
 & \cellcolor{LightCyan} 150$\pm$2 & \cellcolor{LightCyan} 37$\pm$2 & \cellcolor{LightCyan} 850$\pm$42
  & \cellcolor{LightCyan} 1183$\pm$49 & \cellcolor{LightCyan} 170$\pm$10
   & \cellcolor{LightCyan} 253$\pm$15 & \cellcolor{LightCyan} 3406$\pm$161\\
 \hline
 
\multirow{4}{*}{C. Right} & 200 & 87 & 825 & 1205 & 171 & 273 & 3385\\
 & 291 & 177 & 779 & 1151 & 150 & 240 & 3254\\
 & 271 & 164 & 793 & 1108 & 154 & 233 & 3159\\
 & \cellcolor{LightCyan} 254$\pm$39& \cellcolor{LightCyan} 143$\pm$40
 & \cellcolor{LightCyan} 799$\pm$19& \cellcolor{LightCyan} 1155$\pm$40
 & \cellcolor{LightCyan} 158$\pm$9& \cellcolor{LightCyan} 249$\pm$17
 & \cellcolor{LightCyan} 3266$\pm$93\\
 \hline
 
\multirow{4}{*}{Right} & 332 & 225 & 778 & 1051 & 203 & 210 & 3104\\
 & 322 & 211 & 816 & 1072 & 178 & 224 & 3107\\
 & 255 & 145 & 902 & 1256 & 179 & 249 & 3586\\
  & \cellcolor{LightCyan} 303$\pm$34 & \cellcolor{LightCyan} 194$\pm$35
  & \cellcolor{LightCyan} 832$\pm$52 & \cellcolor{LightCyan} 1126$\pm$92
  & \cellcolor{LightCyan} 187$\pm$12 & \cellcolor{LightCyan} 228$\pm$16
  & \cellcolor{LightCyan} 3266$\pm$227\\
 \hline
All & 223$\pm$64 & 111$\pm$65
& 835$\pm$39 & 1173$\pm$63 & 194$\pm$57
& 258$\pm$28 & 3345$\pm$157\\
\hline
\end{tabular}
\label{table:demographic}
\end{table}

\end{document}